\documentclass[12pt]{article}

\usepackage{amsfonts}
\usepackage{amsmath}
\usepackage[dvipdfmx]{hyperref}
\usepackage{cite}
\usepackage{epsfig}
\usepackage{latexsym}
\usepackage{paralist}
\usepackage{fancyhdr}
\usepackage{graphicx}
\numberwithin{equation}{section}
\usepackage[vcentermath]{youngtab}
\usepackage{young}
\usepackage{ytableau}
\usepackage{etex}
\usepackage{braket}
\usepackage{float}

\def\be{\begin{equation}}
\def\ee{\end{equation}}
\def\bea{\begin{eqnarray}}
\def\eea{\end{eqnarray}}

\renewcommand{\thefootnote}{\fnsymbol{footnote}}

\begin{document}

\hfuzz=100pt
\title{{\Large \bf{3d Deconfinement, Product gauge group, Seiberg-Witten and New 3d dualities}}}
\date{}
\author{ Keita Nii$^a$\footnote{keitanii@hri.res.in}
}
\date{\today}

\maketitle

\thispagestyle{fancy}
\cfoot{}
\renewcommand{\headrulewidth}{0.0pt}

\vspace*{-1cm}
\begin{center}
$^{a}${{\it Harish-Chandra Research Institute }}
\\ {{\it Chhatnag Road, Jhusi, Allahabad 211019, India}}

\end{center}

\begin{abstract}
We construct a three dimensional deconfinement method which enables us to find new three-dimensional dualities and we apply various techniques developed in four dimensional supersymmetric gauge theories, such as the product gauge groups and Seiberg-Witten curves to the three dimensional $\mathcal{N}=2$ supersymmetric gauge theories. Dual descriptions of three dimensional $\mathcal{N}=2$ supersymmetric gauge theories which involve two-index matters, for example, adjoint, symmetric, and anti-symmetric matters without superpotentials can be obtained. These matters are described in terms of s-confining phases of the supersymmetric gauge theories. 
\end{abstract}

\renewcommand{\thefootnote}{\arabic{footnote}}
\setcounter{footnote}{0}

\newpage
\tableofcontents 
\clearpage

\section{Introduction}
Supersymmetry is one of the candidates for beyond the Standard Model and attracting much attention as the playground for testing the various ideas of analyzing the strongly-coupled or non-perturbative dynamics, for instance, such as QCD. The supersymmetry severely constrains the perturbative corrections and we can make non-perturbative predictions by combining the holomorphic properties, symmetry argument and the selection rules. About twenty years ago, in a series of analyses of the four-dimensional $\mathcal{N}=1$ supersymmetric gauge theories, Seiberg found a beautiful duality which relates the theories with different ranks of the gauge groups \cite{Seiberg:1994pq}. After that, various dualities which contain not only the (anti-)fundamental matters but also the adjoint, symmetric or anti-symmetric matters were found by many authors (for example, see \cite{Intriligator:1995ax} which includes the comprehensive report on the extension of the Seiberg duality).
 
Among these dualities, the most constructive way of finding dualities is so called ``deconfinement technique'' \cite{Berkooz:1995km,Pouliot:1995me,Luty:1996cg}, where the two-index matters, such as the anti-symmetric matters, are considered as the mesons of some (s-)confining gauge theories. Since the deconfined description only contains the fields in the (anti-)fundamental representations, the usual Seiberg duality \cite{Seiberg:1994pq} can be adapted. Thus the dual description of the $SU(N)$ gauge theories with a two-index matter is given by theories with a product of gauge groups, $SU(N) \times \mbox{(other gauge groups)}$. If the dual description also has the two-index matters, we can again deconfine it by additional gauge groups. In this way, we generally have the infinite duality sequences.

In three space-time dimensions, the dynamics of the supersymmetric gauge theories with four supercharges was analyzed (for example see \cite{Affleck:1982as, Aharony:1997bx, deBoer:1997kr}), where it was found that the phenomena similar to 4d $\mathcal{N}=1$ SQCD arise: For example, chiral symmetry breaking, confinement, dynamical supersymmetry breaking. In addition, 3d SUSY gauge theories have new features, say, Coulomb branch dynamics and real masses, which is closely related to the fact that the 3d vector superfield supplies the Coulomb branch. These new ingredients make the 3d dynamics highly interesting and complicated.   

It is now known that there are various dualities also in 3d: Giveon-Kutasov duality \cite{Giveon:2008zn} relates supersymmetric Chern-Simons matter gauge theories, which is a generalization of the level-ranl duality in the pure Chern-Simons theory. Aharony duality \cite{Aharony:1997gp} is a Seiberg-like duality of the $U(N)$ or $Sp(2N)$ gauge theories without Chern-Simons terms. Furthermore their extensions including adjoint matters are very well studied \cite{Niarchos:2008jb, Niarchos:2009aa}\cite{Kim:2013cma, Park:2013wta}. However the deconfinement technique in 3d have not been investigated yet. 

Recently dynamics of the Coulomb branch in the 3d $\mathcal{N}=2$ SUSY gauge theories have been extensively studied \cite{Intriligator:2013lca, Intriligator:2014fda, Aharony:2013dha}, where it was argued that the effects of the real masses for the matter fields, of the induced Chern-Simons levels and of the Fayet-Iliopoulos terms are important to determine the low-energy dynamics. This understanding of the Coulomb branch dynamics enables us to find the relationship between the 4d Seiberg dualities and the 3d ones \cite{Aharony:2013dha, Aharony:2013kma}. 

For the 3d dualities with two-index matters, we have the 3d version of the Kutasov-Schwimmer duality \cite{Kim:2013cma, Park:2013wta}. The electric side is a $U(N_c)$ gauge theories with $N_f$ (anti-)fundamental matters $Q, \tilde{Q}$ and one adjoint matter $X$ with a superpotential $W= \mathrm{tr}\, X^{k+1}$. The magnetic dual is given by a $U(kN_f-N_c)$ gauge theory. The Kutasov-Schwimmer duality is regarded as a generalization of the conventional Seiberg duality containing only the fundamental matters because the superpotential with some perturbations $W= \sum_{i=1}^{k+1}g_i \mathrm{tr}\, X^i  $ breaks the $U(N_c)$ gauge symmetry into $U(n_1) \times \cdots \times U(n_k) $ and each $U(n_i)$ part has only the fundamental matters. Thus the problem reduces to the conventional Seiberg duality at least in these perturbations. In addition, the superpotential truncates the chiral ring up to $\mathrm{tr} \, X^{k}$ and as the result the moduli space of vacua becomes tractable. To find the duality with two index matters and without any superpotential is extremely non-trivial. In four-dimension we can use the deconfinement technique to find such a duality \cite{Berkooz:1995km,Pouliot:1995me,Luty:1996cg}. However such a technique has not been developed in three-dimensions.

In this paper we will investigate the deconfinement technique for the 3d $\mathcal{N}=2$ supersymmetric theories with two index matters. We will especially focus on the deconfinement of the anti-symmetric matters, which is performed with the s-confining phase of the 3d $\mathcal{N}=2$ $Sp(2N)$ gauge theory with $2N+2$ fundamental matters. However this method is not limited to the anti-symmetrics. It will be easily generalized to the theories with symmetric and adjoint matters. By using the deconfinement technique we will found that large class of the $SU(N)$ gauge theories with one anti-symmetric is completely s-confining. 
When considering the deconfinement we are forced to study the product gauge groups and the structure of their Coulomb branches.  Thus we will consider the $SU(2)^N$ product gauge theory as an illuminating example. This theory was already investigated in 4d \cite{Intriligator:1994sm, Intriligator:1994jr, Csaki:1997zg} and these authors claimed that  we can derive the exact superpotential by using the holomorphy and the so-called ``linearity principle'' (integrating-in method) and that if we include the bi-fundamental matters, the theory is in the Coulomb phase and the $\mathcal{N}=1$ Seiberg-Witten curve can be studied. We will analyze the 3d version by using the quantum dynamics of the 3d $\mathcal{N}=2$ $SU(2)$ gauge theory and we will find the consistency with the 4d exact superpotential and the dimensional reduction  of the Seiberg-Witten curve \cite{Seiberg:1996nz}. 

Organization of this paper is as follows.
In Section 2, we will briefly review the quantum dynamics of the 3d $\mathcal{N}=2$ $Sp(2N)$ and $SU(N)$ gauge theories which will be employed for deconfining the anti-symmetric matters. Depending on the number of (fundamental) flavors, there are various phases.  
In Section 3, we will show the deconfinement of a 3d $\mathcal{N}=2$ $SU(N)$ gauge theory with anti-symmetric matters and (anti-) fundamentals. First example is a chiral gauge theory with one anti-symmetric matter and two fundamentals and $2N$ anti-fundamentals. Next we will study the 3d $\mathcal{N}=2$ $SU(N)$ gauge theory with anti-symmetric matters and vector-like matters in the fundamental representations, which was considered by \cite{Csaki:2014cwa}. 
In Section 4, we will consider the product gauge group $SU(2)^N$.
In Section 5, a 3d $\mathcal{N}=2$ $SU(N)$ gauge theory with one anti-symmetric matter and generic number of the (anti-)fundamentals is considered.
Finally in Section 6 we conclude and discuss the future directions.

\section{Quantum aspects of 3d $\mathcal{N}=2$ $Sp(2N)$ and $SU(N)$ gauge theories}
In this section we will briefly summarize the quantum aspects of the 3d $\mathcal{N}=2$ gauge theories with $SU(N)$ or $Sp(2N)$\footnote{The convention of the $Sp(2N)$ group is that $Sp(2) \sim SU(2)$} gauge symmetry, paying attention to the flat directions (moduli) of the scalar fields, which is based on \cite{Karch:1997ux, Aharony:1997bx}. Since there is a 3d $\mathcal{N}=2$ supersymmetry which means the theory has four supercharges, the theory is obtained from dimensional reduction of the 4d $\mathcal{N}=1$ supersymmetric gauge theory. Thus a 3d chiral superfield contains one complex scalar field as usual. If there is no superpotential and no quantum correction which makes this scalar massive, then the scalar field becomes a flat direction. We will call this flat directions Higgs branch. 

For a vector superfield there is also a real scalar field in the adjoint representation coming from the compactified direction of the 4d gauge field $A_3$ which we denote as $\phi$. We usually diagonalize $\phi$ by the gauge and Weyl transformations 
\begin{align}
\phi=\begin{cases}
    \mathrm{diag}(\phi_1 ,\phi_2 ,\cdots,\phi_N),~~ \phi_1 > \phi_2 >\cdots> \phi_N,~~\sum_{i=1}^{N} \phi_i=0& \mbox{for}~~SU(N)  \\
    \mathrm{diag}( \phi_1 ,\cdots,\phi_N, -\phi_N ,\cdots, -\phi_1),~~\phi_1 > \phi_2 >\cdots> \phi_N & \mbox{for}~~Sp(2N). 
  \end{cases}
\end{align}
When this adjoint scalar $\phi$ takes generic vev $\braket{\phi} \neq 0$, the gauge group is broken to $U(1)^{\mathrm{rank}(G)}$, where $G$ is $SU(N)$ or $Sp(2N)$. Since a three-dimensional photon is dual to a compact scalar field $\sigma$ which we usually call ``dual photon'', combining these two real scalars, each $U(1)$ vector superfield gives one complex scalar field  
\begin{align}
\Phi_i =\phi_i + i \sigma_i.
\end{align}
Thus the vector superfield has the $\mathrm{rank}(G)$ dimensional flat directions at tree-level. However these directions are in many cases lifted by the instanton corrections \cite{Affleck:1982as, Aharony:1997bx, deBoer:1997kr}. In the presence of the fundamental matters, only single branch is un-lifted, which we denote by
\begin{align}
Y \sim\begin{cases}
    \exp(\Phi_1 -\Phi_N) &  \mbox{for}~~SU(N)  \\
    \exp(\Phi_1+\Phi_N) & \mbox{for}~~Sp(2N).
      \end{cases}
\end{align}
This exponential form is due to the fact that the dual photon is actually the compact scalar.
When the above Coulomb branch operators take some vev's, the unbroken $U(1)$ gauge symmetry introduces a topological $U(1)_J$ symmetry whose current is defined as 
\begin{align}
J^\mu_{U(1)_J} =\epsilon^{\mu \nu \rho} F_{\nu \rho}^{U(1)}.
\end{align}
Since this current generates the shift symmetry of the dual photon, the Coulomb branch operator is charged under this global symmetry. When we usually have other $U(1)$ global symmetries, say, $U(1)_B$ which rotates the chiral superfields, the mixed Chern-Simons term between $U(1)_J$ and $U(1)_B$ is induced from the one-loop diagrams and leads to the mixing of the global $U(1)$ symmetries in a non-trivial way. As the result, $Y$ is not charged under the $U(1)_B$ classically but quantum mechanically charged. These mixings are calculated by the Callias index theorem \cite{Callias:1977kg, Weinberg:1979zt, Weinberg:1982ev}. This phenomenon has completely identical origin as the fact that in the Chern-Simons theory the electrically charged particles are propagating with the magnetic flux accompanied.

 Here we will show the global charges of these flat directions and the s-confining phases.
 Let us first consider the 3d $\mathcal{N}=2$ $SU(N)$ gauge theory with $F$ fundamentals $Q$ and $F$ anti-fundamentals $\tilde{Q}$. The global charges of the matter contents and the Coulomb branch operator are as follows, where $\lambda$ is a gaugino.
\begin{table}[H]\caption{Quantum numbers of the electric theory} 
\begin{center}
  \begin{tabular}{|c|c||c|c|c|c|c| } \hline
   &$SU(N)$&$SU(F)_L$& $SU(F)_R$&$U(1)_B$&$U(1)_A$&$U(1)_R$ \\ \hline 
   $Q$&${\tiny \yng(1)}$&${\tiny \yng(1)}$&1&1&1&$R$  \\
 $\tilde{Q}$  &${\tiny \bar{\yng(1)}}$&1&${\tiny \bar{\yng(1)}}$&$-1$&1&$R$ \\
 $\lambda$&$\mathrm{adj.}$&1&1&0&0&1 \\ \hline 
 $Y_{SU(N)}$&1&1&1&0&$-2F$&$-2F(R-1) -2(N-1)$  \\ \hline
  \end{tabular}
  \end{center}\label{qce1}
\end{table}
The monopole-instanton lifts almost all the Coulomb branch and the one complex dimension is expected to be un-lifted. We call it $Y_{SU(N)}$. 
For $F=N$, the theory is in the s-confining phase \cite{Aharony:1997bx}. The effective degrees of freedom are meson $M:=\bar{Q}Q$, baryonic operators $B:=Q^N, \bar{B}:=\bar{Q}^N$ and Coulomb branch operator $Y_{SU(N)}$. The dynamics of the s-confining phase is described by a superpotential
\begin{align}
W=-Y_{SU(N)} (\det M -B \bar{B}),
\end{align}
which is consistent with the global symmetries. 

For a 3d $\mathcal{N}=2$ $Sp(2N)$ theory with $2F$ fundamental matters, the matter contents and the quantum numbers of the Coulomb branch operator are summarized in Table \ref{SP}.
\begin{table}[H]\caption{Quantum numbers of the 3d $\mathcal{N}=2$ $Sp(2N)$ gauge theory} 
\begin{center}
  \begin{tabular}{|c|c||c|c|c| } \hline
   &$Sp(2N)$&$SU(2F)$&$U(1)_B$&$U(1)_R$ \\ \hline 
   $Q$&${\tiny \yng(1)}$&${\tiny \yng(1)}$&1&$R$  \\
 $\lambda$&$\mathrm{adj.}$&1&0&1 \\ \hline 
 $Y_{Sp(2N)}$&1&1&$-2F$&$-2F(R-1) -2N$ \\ \hline
  \end{tabular}
  \end{center}\label{SP}
\end{table}
We expect that the monopole-instantons again lift most of the Coulomb branch and the complex one dimensional part remains un-lifted and it is denoted by $Y_{Sp(2N)}$.  
The s-confining phase arises for $F=N+1$ \cite{Karch:1997ux}, in which the gauge singlets $M:=QQ$ and $Y_{Sp(2N)}$ with a superpotential  
\begin{align}
W=-Y_{Sp(2N)} \mathrm{Pf}\, M.
\end{align}
give the correct description of the low-energy dynamics.

\section{Deconfinement with s-confining $\times$ s-confining}
\subsection{$SU(2N)$}
As an illustration of how the 3d deconfinement technique works, we will first consider 3d $\mathcal{N}=2$ $SU(2N)$ gauge theories with single anti-symmetric matter and some non-vector like (``chiral'') matters in the (anti-)fundamental representations. In 4d, the anti-symmetric matters are deconfined into the (bi-)fundamental quarks by regarding the anti-symmetric matters as the mesons of the s-confining $Sp$ gauge theory \cite{Berkooz:1995km,Pouliot:1995me,Luty:1996cg}. We can apply the same technique to the 3d case more easily as follows. Since we have no chiral anomaly in 3d, without bothering about the chiral anomaly matching between the confined and deconfined theories, we can easily deconfine the anti-symmetric matters by using the s-confining phase of the 3d $\mathcal{N}=2$ $Sp(2N-2)$ theory. Only the subtlety comes from the existence of the Coulomb branch which will be lifted by introducing the additional singlet and by coupling it to the Coulomb branch.  In this subsection, we will consider the case of the $SU(2N)$ gauge group. Next subsection, the $SU(2N+1)$ case will be given.

To be more specific, the first example is a 3d $\mathcal{N}=2$ $SU(2N)$ gauge theory with two fundamentals, $2N$ anti-fundamentals and one anti-symmetric matter. We call this theory ``electric''. The deconfined description is easily obtained by a $SU(2N) \times Sp(2N-2)$ theory. The matter contents and their quantum numbers of the deconfined description are summarized in Table \ref{dec1}. In the table, $Y_{SU(2N)}$ and $Y_{Sp(2N-2)}$ are the Coulomb branch coordinates of the $SU(2N)$ and $Sp(2N-2)$ gauge groups respectively and we here chose the $U(1)_R$ charge as generic values because we do not know the correct $U(1)_R$ charge which is realized in the far infrared region.

\begin{table}[H]\caption{Quantum numbers of the s-confining $\times$ s-confining theory} 
\begin{center}\scalebox{0.78}{
  \begin{tabular}{|c|c|c||c|c|c|c|c|c| } \hline 
 &$SU(2N)$&$Sp(2N-2)$ &$SU(2N)$&  $SU(2)$&$U(1)$&$U(1)$&$U(1)$&$U(1)_R$ \\ \hline 
 $Q$ &$\Box$&$\Box$&$1$&1&1&0&0&$R_1$ \\
$Q''$ &$\Box$&1&1&$\Box$&0&1&0&$R_2$ \\
$\bar{Q}$&$\bar{{\tiny \yng(1)}}$&1&$\Box$&1&0&0&1&$R_3$ \\
S &1&1&1&1&$2N$&0&0&$2NR_1$  \\ \hline 
$Y_{SU(2N)}$&1&1&1&1&$-2N+2$&$-2$&$-2N$&$-(2N-2)R_1-2R_2-2NR_3+2$ \\
$Y_{Sp(2N-2)}$&1&1&1&1&$-2N$&0&0&$-2NR_1+2$ \\ \hline
  \end{tabular}}
  \end{center}\label{dec1}
\end{table}

Due to the special number of the (anti-)fundamentals, both of the gauge groups, $SU(2N)$ and $Sp(2N-2)$ are in the s-confining phases. We call this structure ``s-confining $\times$ s-confining'' according to the 4d case in \cite{Hirayama:1998hu} where the 4d versions of ``s-confining $\times$ s-confining'' were considered. In this theory we can replace both of the gauge dynamics to the descriptions with only the chiral superfields and without gauge fields. The Coulomb branch of the $Sp(2N-2)$ gauge theory is lifted by the singlet field $S$ giving the mass term
\begin{align}
W=SY_{Sp(2N-2)}.
\end{align}

Since the dynamics of the $Sp(2N-2)$ gauge theory is in the s-confining phase, we can first replace the $Sp(2N)$ dynamics to a theory with only the $Sp(2N-2)$-gauge singlets by assuming $g^2_{SU} \ll g^2_{Sp} $. The low-energy effective theory becomes a 3d $\mathcal{N}=2$ $SU(2N)$ gauge theory in Table \ref{CON1}. The matter contents includes the anti-symmetric matter which we wanted to deconfine.

\begin{table}[H]\caption{Quantum numbers of the electric $SU(2N)$ theory} 
\begin{center}\scalebox{1}{
  \begin{tabular}{|c|c||c|c|c|c|c|c| } \hline 
 &$SU(2N)$&$SU(2N)$&  $SU(2)$&$U(1)$&$U(1)$&$U(1)$&$U(1)_R$ \\ \hline  
 $V:=QQ$&${\tiny \yng(1,1)}$&1&1&2&0&0&$2R_1$ \\
$Q''$ &$\Box$&1&$\Box$&0&1&0&$R_2$ \\
$\bar{Q}$ &${\tiny \bar{ \yng(1)}}$&$\Box$&$1$&0&0&1&$R_3$ \\
$Y_{Sp(2N-2)}$ &1&1&1&$-2N$&0&0&$-2NR_1+2$ \\
$S$ &1&1&1&$2N$&0&0&$2NR_1$ \\ \hline
  \end{tabular}}
  \end{center}\label{CON1}
\end{table}

The superpotential describing the s-confining phase of the $Sp(2N-2)$ gauge theory is
\begin{align}
W=-Y_{Sp(2N-2)} V^N +SY_{Sp(2N-2)},
\end{align}
where the notations like $V^N$ is a bit sloppy because more rigorously we should regard this as Pfaffian. But in this article we are satisfied with the above notations for our purposes. Notice that the fields $V$ and $Y_{Sp(2N-2)}$ are not composites but elementary fields from the low-energy $SU(2N)$ gauge theory point of view since we are in the s-confining phase of the $Sp(2N)$ theory.
Due to the mass term between $S$ and $Y_{Sp(2N-2)}$, the first term plays no role at all and the equation of motion for $Y_{Sp(2N-2)}$ gives $S= V^N$. At the low-energy limit, we have the 3d $\mathcal{N}=2$ $SU(2N)$ gauge theory with two fundamentals, $2N$ anti-fundamentals and one anti-symmetric matter with no tree-level superpotential.

The Coulomb branch of this theory is complicated due to the reason discussed in \cite{Csaki:2014cwa, Amariti:2015kha}, where it was argued that the presence of the anti-symmetric matter and the ``chiral'' nature of the (anti-) fundamental matters make the Coulomb branch complicated since the various mixed Chern-Simons terms and the FI terms are effectively generated along the Coulomb branch and this leads to the modification of the monopole-instanton structure and making the Coulomb branch non gauge invariant. This may allow us to have the two-dimensional Coulomb branch. Then in this case we also expect more than one coordinate to be required for describing the entire Coulomb branch. The Coulomb branch direction
\begin{align}
Y \sim \left(
    \begin{array}{ccccc}
\sigma &&&& \\
&0& && \\
&&\ddots&& \\
&&&0& \\
&&&&-\sigma
    \end{array}
  \right)
\end{align}
which is usually globally defined and un-lifted should be dressed by some matter fields. This is because the chirality of the matter contents induces the effective Chern-Simons term along the direction $Y$, and then $Y$ becomes non-gauge invariant. The dressed Coulomb branch operators which are expected to be un-lifted are as follows.
\begin{align}
YV^{N-1},~~YV^{N-2} Q''^2 
\end{align}
The global charges of these operators are in Table \ref{Dress}.

\begin{table}[H]\caption{Quantum numbers of the (dressed) monopole operators} 
\begin{center}\scalebox{0.96}{
  \begin{tabular}{|c||c|c|c|c|c|c| } \hline 
  &$SU(2N)$&$SU(2)$&$U(1)$&$U(1)$&$U(1)$&$U(1)_R$ \\ \hline
$Y$&1&1&$-4N+4$&$-2$&$-2N$&$-(4N-4)R_1-2R_2-2NR_3+2 $\\ 
$YV^{N-1}$&1&1&$-2N+2$&$-2$&$-2N$&$-(2N-2)R_1-2R_2-2NR_3+2$  \\
$YV^{N-2} Q''^2 $&1&1&$-2N$&0&$-2N$& $-2NR_1-2NR_3+2$ \\ \hline
  \end{tabular}}
  \end{center}\label{Dress}
\end{table}

Here we will briefly show that these composite operators are indeed gauge invariant according to the discussion in \cite{Csaki:2014cwa, Amariti:2015kha}. First, along the Coulomb branch direction $Y$, the gauge group is broken as 
\begin{align}
SU(2N) \rightarrow SU(2N-2) \times U(1)_1 \times U(1)_2,
\end{align}
where the indices of the $U(1)_1$ and $U(1)_2$ are only labeling the each $U(1)$ and do not indicate the Chern-Simons level. The generators of these two $U(1)$'s are conveniently chosen as 
\begin{align}
T_{U(1)_1} = \left(
    \begin{array}{ccccc}
1 &&&& \\
&0& && \\
&&\ddots&& \\
&&&0& \\
&&&&-1
    \end{array}
  \right),~~~T_{U(1)_2} = \left(
    \begin{array}{ccccc}
N-1 &&&& \\
&-1& && \\
&&\ddots&& \\
&&&-1& \\
&&&&N-1
    \end{array}
  \right),
\end{align}
where we neglect the normalization which is not important here.
Under this gauge symmetry breaking, the matter fields decomposed as
\begin{align}
\Box &\rightarrow \Box_{(0,-1)} +1_{(1,N-1)} +1_{(-1,N-1)} \\
\bar{{\tiny \yng(1)}} &\rightarrow  \bar{{\tiny \yng(1)}}_{(0,1)} +1_{(-1,1-N)} +1_{(1,1-N)} \\
{\tiny \yng(1,1)} & \rightarrow {\tiny \yng(1,1)}_{(0,-2)} +\Box_{(1,N-2)} +\Box_{(-1,N-2)} +1_{(0,2N-2)}.
\end{align}

Along the $Y$ direction, some of the above fields are massive and should be integrated out. Then the mixed Chern-Simons term is induced for $U(1)_1$ and $U(1)_2$:
\begin{align}
k_{eff}^{U(1)_1,U(1)_2} =-2N+2.
\end{align}
This mixed Chern-Simons term makes the Coulomb branch operator $Y$ gauge non-invariant under the $U(1)_2$ gauge group. The $U(1)_2$ charge of $Y$ is $2N-2$. Then we can make this operator gauge invariant by multiplying a field in a representation ${\tiny \bar{\yng(1,1)}}$ because the ${\tiny \bar{\yng(1,1)}}$ field decomposes under the above gauge symmetry breaking as
\begin{align}
{\tiny \bar{\yng(1,1)}} \rightarrow {\tiny \bar{\yng(1,1)}}_{(0,2) } +\bar{\Box}_{-1,2-N} +\bar{\Box}_{1,2-N} +1_{(0,2-2N)}, 
\end{align}
and we can use the last component $1_{(0,2-2N)}$ for making the Coulomb branch operator gauge invariant.
Thus we have two types of the dressed monopole operators:
\begin{align}
YV^{N-1},~~YV^{N-2} Q''^2.
\end{align}
%

Next we will replace the $SU(2N)$ dynamics to the s-confining one by assuming $g^2_{Sp(2N-2)} \ll g^2_{SU(2N)}$. The theory becomes a 3d $\mathcal{N}=2$ $Sp(2N)$ gauge theory with the following matter contents. We call this theory ``magnetic''.
\begin{table}[H]\caption{Quantum numbers of the magnetic $Sp(2N)$ gauge theory} 
\begin{center}\scalebox{0.81}{
  \begin{tabular}{|c|c||c|c|c|c|c|c| } \hline 
 &$Sp(2N-2)$&$SU(2N)$&  $SU(2)$&$U(1)$&$U(1)$&$U(1)$&$U(1)_R$ \\ \hline  
$M:=Q \bar{Q}$ &$\Box$&$\Box$&1&1&0&1&$R_1+R_3$ \\
$M_1:=Q'' \bar{Q}$&1&$\Box$&$\Box$&0&1&1&$R_2+R_3$ \\
$B:= Q^{2N-2} Q''^2$&1&1&1&$2N-2$&2&0&$(2N-2)R_1 +2R_2$ \\
$\bar{B}:= \bar{Q}^{2N}$&1&1&1&0&0&$2N$&$2NR_3$ \\
$Y_{SU(2N)}$&1&1&1&$-2N+2$&$-2$&$-2N$& $-(2N-2)R_1 -2R_2-2NR_3 +2$ \\ 
$S$&1&1&1&$2N$&0&0&$2NR_1$ \\ \hline
$Y_{Sp(2N-2)}^{\mathrm{mag}}$&1&1&1&$-2N$&0&$-2N$& $-2NR_1-2NR_3+2$ \\ \hline
  \end{tabular}}
  \end{center}\label{qcml1}
\end{table}

In the deconfinement description with the $SU(2N) \times Sp(2N-2)$ gauge symmetry, the $Sp$-monopole has the fermionic zero modes coming from $Q$.
Thus this $Sp$-monopole (instanton) should behave as the vertex for $Q$.
On the other hand, in the s-confined description with the $Sp(2N-2)$ gauge symmetry, we have only the $SU(2N)$ singlets and thus the $Sp(2N)$-monopole has the fermionic zero-modes from the mesino $M:=Q \bar{Q}$.
As the result, these two $Sp(2N)$-monopoles have the different types of the fermionic zero-modes although the numbers of the zero-modes are identical. Then, a naive relation $Y_{Sp(2N-2)} \sim Y_{Sp(2N-2)}^{\mathrm{mag}} $ is not correct. This fact can be confirmed from the quantum numbers of these operators. More correctly the operator matching of the Coulomb branch coordinates of the $Sp(2N-2)$ gauge groups is
\begin{align}
Y_{Sp(2N-2)} = Y_{Sp(2N-2)}^{\mathrm{mag}} \bar{B}
\end{align}
up to an unimportant normalization factor.
The role of $\bar{B}$ is expected to absorb the unwanted zero-modes coming from the $\bar{Q}$ superfields. Thus the mass term for $Y_{Sp(2N-2)} $ is mapped into the Yukawa term
\begin{align}
W=S Y_{Sp(2N-2)}^{\mathrm{mag}} \bar{B}
\end{align}
and the s-confining phase of the $SU(2N)$ gauge theory is described by 
\begin{align}
W=-Y_{SU(2N)} (M^{2N-2} M_1^2 -B \bar{B}) +SY_{Sp(2N-2)}^{\mathrm{mag}} \bar{B}.
\end{align}

The resulting theory is again s-confining since the magnetic $Sp(2N-2)$ theory has $2N$ fundamentals $M$, thus we move on to the description only with the gauge singlets. The matter contents and their global charges are summarized in the following table.

\begin{table}[H]\caption{Quantum numbers of the non-gauge theory} 
\begin{center}\scalebox{0.97}{
  \begin{tabular}{|c||c|c|c|c|c|c| } \hline 
  &$SU(2N)$&  $SU(2)$&$U(1)$&$U(1)$&$U(1)$&$U(1)_R$ \\ \hline  
  $N:=MM$&${\tiny  \yng(1,1) }$&1&2&0&2&$2R_1+2R_3$  \\
 $M_1 $ &$\Box$&$\Box$&0&1&1& $R_2+R_3$ \\
 $B$ &1&1&$2N-2$&2&0&$(2N-2)R_1 +2R_2$    \\
$\bar{B}$ &1&1&0&0&$2N$&$2NR_3$ \\
$Y_{SU(2N)}$&1&1&$-2N+2$&$-2$&$-2N$&$-(2N-2)R_1 -2R_2-2NR_3 +2$ \\
$Y_{Sp(2N-2)}^{\mathrm{mag}}$&1&1&$-2N$&0&$-2N$& $-2NR_1-2NR_3+2$  \\
$S$&1&1&$2N$&0&0& $2NR_1$ \\ \hline
  \end{tabular}}  
  \end{center}\label{qcml1}
\end{table}
The superpotential describing the low-energy dynamics is
\begin{align}
W=-Y_{Sp(2N-2)}^{\mathrm{mag}} N^N -Y_{SU(2N)}(N^{N-1}M_1^2 -B \bar{B} ) +SY_{Sp(2N-2)}^{\mathrm{mag}}  \bar{B}   \label{scnsup}
\end{align}
and the operator matching between the electric and magnetic sides are
\begin{gather}
S=V^N,~~B=V^{N-1}Q''^2,~~N=V\bar{Q}^2,~~M_1=Q''\bar{Q},~~\bar{B}=\bar{Q}^{2N}   \nonumber \\
Y_{SU(2N)} =YV^{N-1},~~Y_{Sp(2N-2)}^{\mathrm{mag}}=YV^{N-2}Q''^2.
\end{gather}
In this way we can find that the 3d $\mathcal{N}=2$ $SU(2N)$ gauge theory with two fundamental matters, $2N$ anti-fundamental matters and one anti-symmetric matter is a s-confining theory. By introducing the mass terms for the (anti-)fundamentals, we can easily flow to the theories with lower numbers of (anti-)fundamentals.

As the check of the above duality and the validity of the 3d deconfinement method, we can derive this s-confining description from 4d. We know that the 4d $\mathcal{N}=1$ $SU(2N)$ gauge theory with one anti-symmetric matter $V$, $2N$ anti-fundamental matters $\bar{Q}$ and $4$ fundamental matters $Q$ is s-confining \cite{Murayama:1995ng,Pouliot:1995me,Poppitz:1995fh,Csaki:1996sm,Csaki:1996zb,Csaki:1998th}. The s-confined (magnetic) description is given by the following chiral superfields,
\begin{table}[H]\caption{Quantum numbers of the non-gauge theory} 
\begin{center}\scalebox{1}{
  \begin{tabular}{|c|c||c|c|} \hline 
&$SU(2N)$&$SU(2N)$&$SU(4)$  \\  \hline
$V$&${\tiny  \yng(1,1) }$&1&1  \\
$\bar{Q}$ &${\tiny  \bar{\yng(1) }}$&${\tiny  \yng(1) }$&1 \\
$Q$&${\tiny  \yng(1) }$&1&${\tiny  \yng(1) }$  \\   \hline
$M:=Q \bar{Q}$&1&${\tiny  \yng(1) }$&${\tiny  \yng(1) }$  \\
$\bar{H}:=A\bar{Q}^2$&1&${\tiny  \yng(1,1) }$&1  \\  
$T:=V^N$&1&1&1  \\
$H_1:=V^{N-1} Q^2$&1&1&${\tiny  \yng(1,1) }$  \\
$H_2:= V^{N-2}Q^4$&1&1& 1 \\
$\bar{B}:=\bar{Q}^{2N}$&1&1&1  \\ \hline
  \end{tabular}}  
  \end{center}\label{qcml1}
\end{table}
\noindent and the superpotential
\begin{align}
W= TM^4 \bar{H}^{N-2} +H_1 M^2 \bar{H}^{N-1}+ H_2 \bar{H}^N +\bar{B} T  H_2+\bar{B} H_1^2,
\end{align}
where we omitted the scaling factor for simplicity. The 3d $\mathcal{N}=2$ SUSY theories and their dualities are obtained by the dimensional reduction of the 4d $\mathcal{N}=1$ dualities \cite{Aharony:2013dha,Aharony:2013kma}. 

First we put the electric and magnetic theories on a circle and introduce the non-perturbative superpotential coming from the Kaluza-Klein monopoles, which is absent in the s-confining magnetic description. In order to switch off the KK-monopole contribution, we need to turn on the real masses for the fundamental matters by gauging the subgroup of the $SU(4)$ global symmetry, whose generator is 
\begin{align}
T= \left(
    \begin{array}{cccc}
     0&&& \\
     &0&&  \\
     &&1&  \\
     &&&-1
    \end{array}
  \right).
\end{align}
This background gauging introduces the real masses for the last two components of $Q$. If we take the low-energy limit, the superpotential induced by the KK-monopole vanishes. Then the electric theory reduces to the 3d $\mathcal{N}=2$ $SU(2N)$ gauge theory with one anti-symmetric $V$, $2N$ anti-fundamentals and two fundamentals $Q$ without superpotentials. On the magnetic side, following fields are only massless:
\begin{gather}
\hat{M}:=M_{a,i} ~~~(a=1,\cdots,2N, ~i=1,2 ) \nonumber \\
\bar{H}, ~~T,~~ H_2,~~\bar{B}  \nonumber \\
h_1:=H_{1}^{12},~~h_2:=H_{1}^{34}
\end{gather}
The low-energy superpotential becomes
\begin{align}
W=h_2  \hat{M}^2  \bar{H}^{N-1}  + H_2 \bar{H}^N +\bar{B}T H_2 + \bar{B}h_1h_2,
\end{align}
which is precisely the same as \eqref{scnsup} if we notice the identification
\begin{gather}
\hat{M}  \sim  M_1 ,~~\bar{H}\sim N, ~~T \sim S \\
h_1 \sim B,~~ h_2 \sim Y_{SU(2N)}, H_2 \sim Y_{Sp(2N-2)}^{\mathrm{mag}}.
\end{gather}

We could alternatively start with the following deconfined theory with the ``s-confining $\times$ s-confining'' structure (Table \ref{ALT1}).
In this case we also have the same result except for the additional singlet to be added, then we will briefly show the result.

\begin{table}[H]\caption{Quantum numbers of the deconfined theory with s-confining $\times$ s-confining structure} 
\begin{center}\scalebox{0.9}{
  \begin{tabular}{|c|c|c||c|c|c|c|c|c| } \hline 
  & $SU(2N)$ & $Sp(2N)$   &$SU(2)$&  $SU(2N)$&$U(1)$&$U(1)$&$U(1)$&$U(1)_R$ \\ \hline
  $Q$ & $\Box$&$\Box$ &1& 1 &1&0&0 & $R_1$ \\
$Q'$&1&$\Box$&$\Box$&1 &0&1&0& $R_2$\\
$\bar{Q}$&$\bar{{\tiny \yng(1) }}$&1&1&$\Box$ &0&0&1&  $R_3$\\ \hline
$Y_{SU(2N)}$&1&1&1&1 &$-2N$&$0$&$-2N$&$-2NR_1-2NR_3 +2$ \\
$Y_{Sp(2N)}$&1&1&1&1 &$-2N$&$-2$&0 & $-2NR_1 -2R_2 +2$ \\ \hline
  \end{tabular}}
  \end{center}\label{ALT1}
\end{table}

We find that both parts of the $SU(2N)$ gauge group and of the $Sp(2N)$ are in the s-confining phases.
First we consider the case with $g^2_{Sp} \gg g^2_{SU}$ where the $Sp(2N)$ gauge theory first confines:

 \begin{table}[H]\caption{Quantum numbers of the electric theory} 
\begin{center}\scalebox{1}{
  \begin{tabular}{|c|c||c|c|c|c|c|c| } \hline 
  & $SU(2N)$ & $SU(2)$&  $SU(2N)$&$U(1)$&$U(1)$&$U(1)$&$U(1)_R$ \\ \hline
 $V:=QQ$&${\tiny \yng(1,1)}$&1&1&2&0&0& $2R_1$ \\
 $V':=QQ'$&$\Box$&$\Box$&1&1&1&0& $R_1+R_2$\\
 $V'':=Q'Q'$ &1&1&1&0&2&0& $2R_2$ \\
 $\bar{Q}$&$\bar{{\tiny \yng(1) }}$&1&$\Box$&0&0&1&  $R_3$ \\
 $Y_{Sp(2N)}$&1&1&1&$-2N$&$-2$&0& $-2NR_1 -2R_2+2$ \\ \hline 
  \end{tabular}}
  \end{center}\label{qcml1}
\end{table}

We call this theory electric and the theory is a 3d $\mathcal{N}=2$ $SU(2N)$ gauge theory with chiral (anti-) fundamental matters and an anti-symmetric matters. The superpotential is
\begin{align}
W=- Y_{Sp(2N)}(V^N V'' +V^{N-1} V'^2).
\end{align}
If we introduce the singlet field $S$ coupling to the Coulomb branch of the $Sp(2N)$ vector superfield
\begin{align}
W=S Y_{Sp(2N)},
\end{align}
the fields $S$, $Y_{Sp}$ are massive and integrated out, the field $V''$ is completely decouples from the other sectors. As the result, we obtain the 3d $\mathcal{N}=2$ $SU(2N)$ gauge theory with two fundamental matters, $2N$ anti-fundamentals and one anti-symmetric matter without superpotentials in the infrared limit.

Next we will consider $g^2_{Sp} \ll g^2_{SU}$, where we can replace the dynamics of the $SU(2N)$ part with the s-confining one, in which the theory only contains the fundamentals and the gauge singlets. We call this description ``magnetic theory''. The quantum numbers of the magnetic theory are given in Table \ref{ALTM}.

\begin{table}[H]\caption{Quantum numbers of the magnetic theory} 
\begin{center}\scalebox{0.97}{
  \begin{tabular}{|c|c||c|c|c|c|c|c| } \hline 
  & $Sp(2N)$ & $SU(2)$&  $SU(2N)$&$U(1)$&$U(1)$&$U(1)$&$U(1)_R$ \\ \hline
 $M:=Q \bar{Q}$&$\Box$&1&$\Box$&1&0&1& $R_1+R_3$ \\
 $B:=Q^{2N}$&$1$&$1$&1&$2N$&0&0& $2NR_1 $ \\
 $\bar{B} :=\bar{Q}^{2N}$ &1&1&1&0&0&$2N$ & $2NR_3$\\
 $Q'$&$\Box$&$\Box$&$1$&0&1&0& $R_2$\\
 $Y_{SU(2N)}$&1&1&1&$-2N$&$0$&$-2N$& $-2N R_1 -2N R_3 +2$ \\ \hline  
 $Y_{Sp(2N)}^{\mathrm{mag}}$&1&1&1&$-2N$&$-2$&$-2N$& $-2N R_1 -2N R_3 -2R_2 +2$  \\ \hline
  \end{tabular}}
  \end{center}\label{ALTM}
\end{table}

The relation between the original $Y_{Sp(2N)}$ and $Y_{Sp(2N)}^{\mathrm{mag}}$ can be found from a symmetry argument as
\begin{align}
Y_{Sp(2N)} = Y_{Sp(2N)}^{\mathrm{mag}} \bar{B},
\end{align}
so the mass term for $Y_{Sp(2N)} $ is mapped into
\begin{align}
W=S Y_{Sp(2N)}^{\mathrm{mag}} \bar{B}
\end{align}
and the superpotential which describes the s-confining phase is 
\begin{align}
W=-Y_{SU(2N)} (M^{2N}  -B \bar{B}).
\end{align}

The theory is a 3d $\mathcal{N}=2$ $Sp(2N)$ gauge theory with the $2N+2$ fundamental flavors $M, Q'$ and the gauge singlets $B, \bar{B}$ and $Y_{SU(2N)}$. Notice that the magnetic $Sp(2N)$ gauge theory is again s-confining, then we have the alternative description with only the gauge singlets (Table \ref{ALTM2}).

\begin{table}[H]\caption{Quantum numbers of the s-confining description of the magnetic theory} 
\begin{center}\scalebox{1}{
  \begin{tabular}{|c|c|c|c|c|c|c| } \hline 
   &$SU(2)$&  $SU(2N)$&$U(1)$&$U(1)$&$U(1)$&$U(1)_R$ \\ \hline 
   $B:=Q^{2N}$&1&1&$2N$&0&0& $2NR_1 $ \\
   $\bar{B}:=\bar{Q}^{2N}$&1&1&0&0&$$2N& $2NR_3$ \\
   $Y_{SU(2N)}$&1&1&$-2N$&$0$&$-2N$& $-2NR_1 -2NR_3 +2$ \\
   $N_1:=MM$&1&${\tiny \yng(1,1)}$&2&0&2& $2R_1+2R_3$ \\
   $N_2:=MQ'$&$\Box$&$\Box$&1&1&1& $R_1+R_2+R_3$ \\
   $N_3:=Q'Q'$&1&1&0&2&0& $2R_2$ \\
   $Y^{\mathrm{mag}}_{Sp(2N)}$&1&1&$-2N$&$-2$&$-2N$& $-2NR_1 -2NR_3-2R_2+2$ \\ \hline
  \end{tabular}}
  \end{center}\label{ALTM2}
\end{table}
The superpotential is rewritten as follows:
\begin{align}
W &=-Y_{SU(2N)} (M^{2N} -B\bar{B}) -Y_{Sp(2N)}^{\mathrm{mag}} (N_1^N N_3 +N_1^{N-1} N_2^2) +S Y_{Sp(2N)}^{\mathrm{mag}} \bar{B}\nonumber \\
&\rightarrow -Y_{SU(2N)} (N_1^N -B \bar{B}) -Y_{Sp(2N)}^{\mathrm{mag}} (N_1^N N_3 +N_1^{N-1} N_2^2)+S Y_{Sp(2N)}^{\mathrm{mag}} \bar{B}.
\end{align}
$N_3$ is identified with $V''$ in the electric theory.

\subsection{$SU(2N+1)$}
For the $SU(2N+1)$ gauge group, we start with the following s-cofining $\times$ s-confining theory (Table \ref{ODDSS1}) with a $\mathcal{N}=2$ supersymmetry. We need an additional $Sp(2N)$ fundamental matter in order to have the s-cofining $\times$ s-confining structure.
\begin{table}[H]\caption{Quantum numbers of the s-confining $\times$ s-confining theory for $SU(2N+1)$} 
\begin{center}
\scalebox{0.74}{
  \begin{tabular}{|c|c|c||c|c|c|c|c|c| } \hline 
 &$SU(2N+1)$&$Sp(2N)$ &$SU(2N+1)$&  $U(1)$&$U(1)$&$U(1)$&$U(1)$&$U(1)_R$ \\ \hline 
 $Q$ &$\Box$&$\Box$&$1$&1&0&0&0&$R_1$ \\
 $Q'$&1&$\Box$&1&0&1&0&0&$R_2$ \\
$Q''$ &$\Box$&1&1&0&0&1&0&$R_3$ \\
$\bar{Q}$&$\bar{{\tiny \yng(1)}}$&1&$\Box$&0&0&0&1&$R_4$ \\
$S$ &1&1&1&$2N+1$&1&0&0&$(2N+1)R_1+R_2$  \\ \hline 
$Y_{SU(2N+1)}$&1&1&1&$-2N$&$0$&$-1$&$-2N-1$&$-2NR_1-R_3-(2N+1)R_4+2$ \\
$Y_{Sp(2N)}$&1&1&1&$-2N-1$&$-1$&0&0&$-(2N+1)R_1-R_2+2$ \\ \hline
  \end{tabular}}
  \end{center}\label{ODDSS1}
\end{table}
The Coulomb branch for the $Sp(2N)$ gauge group is lifted by introducing the mass term just as before:
\begin{align}
W=SY_{Sp(2N)}.
\end{align}
Replacing the dynamics of the $Sp(2N)$ gauge theory with the s-confining description, we have the 3d $\mathcal{N}=2$ $SU(2N+1)$ gauge theory with two fundamental matters, $2N+1$ anti-fundamental matters and one anti-symmetric matter. The matter contents and their global charges are summarized in the following table.

\begin{table}[H]\caption{Quantum numbers of the $SU(2N+1)$ ``electric'' gauge theory} 
\begin{center}\scalebox{0.89}{
  \begin{tabular}{|c|c||c|c|c|c|c|c| } \hline 
 &$SU(2N+1)$ &$SU(2N+1)$&  $U(1)$&$U(1)$&$U(1)$&$U(1)$&$U(1)_R$ \\ \hline 
$V:=QQ$ &${\tiny \yng(1,1)}$&1&2&0&0&0&$2R_1$ \\
$V':=QQ'$&$\Box$&1&1&1&0&0&$R_1+R_2$  \\
$Q''$ &$\Box$&1&0&0&1&0&$R_3$   \\
$\bar{Q}$&${\tiny \bar{\yng(1)} }$&$\Box$&0&0&0&1&$R_4$ \\
$Y_{Sp(2N)}$&1&1&$-2N-1$&$-1$&0&0&$-(2N+1)R_1 -R_2+2$ \\ 
$S$ &1&1&$2N+1$&1&0&0&$(2N+1)R_1+R_2$\\ \hline
   \end{tabular}}
  \end{center}\label{oddele1}
\end{table}
The superpotential describing the above s-confining phase is 
\begin{align}
W=-Y_{Sp(2N)} V^N V' +SY_{Sp(2N)}.
\end{align}
If we use the equation of motion of $S$, the first term of the superpotential palys no role.
Then, at the low-energy limit, we have the 3d $\mathcal{N}=2$ $SU(2N+1)$ gauge theory with the two fundamentals, the $2N+1$ anti-fundamentals and the one anti-symmetric matter and without any tree-level superpotential.
 
If we consider the case with $g^2_{SU} \gg g^2_{Sp}$, we first change the $SU(2N+1)$ dynamics to the s-confining one. The theory becomes a 3d $\mathcal{N}=2$ $Sp(2N)$ gauge theory with the following field contents and we call this theory ``magnetic'':
\begin{table}[H]\caption{Quantum numbers of the $Sp(2N)$ ``magnetic'' gauge theory}
\begin{center} 
\scalebox{0.77}{
  \begin{tabular}{|c|c||c|c|c|c|c|c| } \hline 
 &$Sp(2N)$ &$SU(2N+1)$&  $U(1)$&$U(1)$&$U(1)$&$U(1)$&$U(1)_R$ \\ \hline 
$M:=Q \bar{Q}$&$\Box$&$\Box$&1&0&0&1&$R_1+R_4$ \\
$M_1:=Q'' \bar{Q}$&1&$\Box$&0&0&1&1&$R_3+R_4$ \\
$B:=Q^{2N}Q''$&1&1&$2N$&0&1&0&$2NR_1+R_3$ \\
$\bar{B}:=\bar{Q}^{2N+1}$&1&1&0&0&0&$2N+1$&$(2N+1)R_4$ \\
$Q'$&$\Box$&1&0&1&0&0&$R_2$ \\
$Y_{SU(2N+1)}$&1&1&$-2N$&$0$&$-1$&$-2N-1$&$-2NR_1-R_3-(2N+1)R_4+2$ \\ 
$S$ &1&1&$2N+1$&1&0&0&$(2N+1)R_1+R_2$\\  \hline
$Y_{Sp(2N)}^{\mathrm{mag}}$&1&1&$-2N-1$&$-1$&$0$&$-2N-1$&$-(2N+1)R_1-(2N+1)R_4 -R_2+2$  \\ \hline
   \end{tabular}}
  \end{center}\label{qcml1}
\end{table}
The Coulomb branch coordinates on the electric and magnetic sides are identified as
\begin{align}
Y_{Sp(2N)} =Y_{Sp(2N)}^{\mathrm{mag}} \bar{B},
\end{align}
and the superpotential describing the s-confining phase of the part of the $SU(2N+1)$ gauge theory is 
\begin{align}
W=-Y_{SU(2N+1)}(M^{2N} M_1 -B \bar{B} ) +SY_{Sp(2N)}^{\mathrm{mag}} \bar{B}.
\end{align}
Notice that this theory is again s-confining, so we can go on to the non-gauge theory:

\begin{table}[H]\caption{Quantum numbers of the $Sp(2N)$ gauge theory} 
\begin{center}\scalebox{0.85}{
  \begin{tabular}{|c||c|c|c|c|c|c| } \hline 
 &$SU(2N+1)$&  $U(1)$&$U(1)$&$U(1)$&$U(1)$&$U(1)_R$ \\ \hline 
$N:=MM$&${\tiny \yng(1,1)}$&2&0&0&2&$2R_1+2R_4$ \\
$N_1:=MQ'$&$\Box$&1&1&0&1&$R_1+R_2+R_4$ \\
$M_1$&$\Box$&0&0&1&1&$R_3+R_4$  \\
$B$&1&$2N$&0&1&0&$2NR_1+R_3$  \\
$\bar{B}$&1&0&0&0&$2N+1$&$(2N+1)R_4$  \\
$Y_{SU(2N+1)}$&1&$-2N$&0&$-1$&$-2N-1$&$-2NR_1-R_3-(2N+1)R_4+2$ \\
$Y_{Sp(2N)}^{\mathrm{mag}}$&1&$-2N-1$&$-1$&0&$-2N-1$&$-(2N+1)R_1-(2N+1)R_4 -R_2+2$ \\
$S$&1&$2N+1$&1&0&0&$(2N+1)R_1+R_2$ \\ \hline
   \end{tabular}}
  \end{center}\label{qcml1}
\end{table}
The superpotential becomes
\begin{align}
W=-Y_{SU(2N+1)} (N^N M_1-B \bar{B}) -Y_{Sp(2N)}^{\mathrm{mag}}N^{N} N_1 +SY_{Sp(2N)}^{\mathrm{mag}} \bar{B}.
\end{align}
The operator matching to the theory in Table \ref{oddele1} can be easily found from the global symmetries:
\begin{gather}
N  = V \bar{Q}^2, ~~N_1= V'\bar{Q} ,~~M_1=Q''\bar{Q} \nonumber \\
B=V^N Q'',~~ \bar{B}=\bar{Q}^{2N+1},~~S=V^NV'.
\end{gather}

\subsection{$SU(2N)$ with an anti-symmetric flavor}
In the previous subsections, we have considered the deconfinement of the $SU(N)$ gauge theory with one anti-symmetric matter and ``chiral'' (anti-)fundamental matters, where the anti-symmetric matters were described in terms of the s-confining theory with the $Sp$ gauge groups. Due to the special number of the (anti-)fundamentals, the part of the $SU(N)$ gauge groups is also s-confining. According to the 4d case in \cite{Hirayama:1998hu}, we call this structure ``s-confining $\times$ s-confining''. In this section, we will further study the cases with ``s-confining $\times$ s-confining''. We will consider $SU(2N)$ gauge theories with matters in representations of ${\tiny \yng(1,1)}$ and ${\tiny \bar{\yng(1,1)}}$.

The first example is a 3d $\mathcal{N}=2$ $SU(4)$ gauge theory with $2 ({\tiny \yng(1)}+\bar{{\tiny \yng(1)}}) $and $2~ {\tiny \yng(1,1)}$, where the fields ${\tiny \yng(1,1)}$ and ${\tiny \bar{\yng(1,1)}}$ in a $SU(4)$ gauge group are in the same representation. This theory was first considered by \cite{Csaki:2014cwa} where it was concluded by the dimensional reduction method from 4d that the theory is s-confining. The global charges of the matter contents are chosen as follows (Table \ref{C1}) for coincidence with ones in \cite{Csaki:2014cwa}. We call this theory ``electric''. 

\begin{table}[H]\caption{Quantum numbers of the electric theory} 
\begin{center}
  \begin{tabular}{|c|c||c|c|c|c|c|c|c| } \hline
  &$SU(4)$&$SU(2)_L$&$SU(2)_R$&$SU(2)_A$&$U(1)$&$U(1)$&$U(1)$&$U(1)_R$ \\ \hline
  $A$&${\tiny \yng(1,1)}$&1&1&$\Box$&0&0&$-3$&0 \\
$Q$  &${\tiny \yng(1)}$&$\Box$&1&1&1&0&2&$\frac{1}{3}$ \\
$\bar{Q}$&${\tiny \bar{\yng(1)}}$&1&$\Box$&1&0&$-1$&$2$&$\frac{1}{3}$ \\ \hline
  \end{tabular}
  \end{center}\label{C1}
\end{table}

This theory has the following deconfinement description in Table \ref{C2} with the $SU(4) \times SU(2) \times SU(2)$ gauge symmetry with a superpotential
\begin{align}
W=S_1 Y_{SU(2)_1} +S_2  Y_{SU(2)_2},
\end{align}
which lifts all the $SU(2) \times SU(2)$ Coulomb branches.

\begin{table}[H]\caption{Quantum numbers of the deconfinement theory} 
\begin{center}\scalebox{0.96}{
  \begin{tabular}{|c|c|c|c||c|c|c|c|c|c|c| } \hline
  &$SU(4)$&$SU(2)_1$&$SU(2)_2$&$SU(2)_L$&$SU(2)_R$&$U(1)$&$U(1)$&$U(1)$&$U(1)'$&$U(1)_R$\\ \hline
$x$&${\tiny \yng(1)}$&${\tiny \yng(1)}$&1&1&1&0&0&$-\frac{3}{2}$&1&$0$ \\
$\bar{x}$&${\tiny \bar{\yng(1)}}$&1&${\tiny \yng(1)}$&1&1&0&0&$-\frac{3}{2}$&$-1$&0 \\
$Q$&${\tiny \yng(1)}$&1&1&${\tiny \yng(1)}$&1&1&0&2&0&$\frac{1}{3}$ \\
$\bar{Q}$&${\tiny \bar{\yng(1)}}$&1&1&1&${\tiny \yng(1)}$&$0$&$-1$&$2$&0&$\frac{1}{3}$ \\ 
$S_1$&1&1&1&1&1&0&0&$-6$&$4$&$0$ \\
$S_2$&1&1&1&1&1&0&0&$-6$&$-4$&0 \\ \hline
$Y_{SU(4)}$&1&1&1&1&1&$-2$&2&$-2$&0&$\frac{2}{3}$ \\
$Y_{SU(2)_1}$&1&1&1&1&1&0&0&6&$-4$&$2$ \\
$Y_{SU(2)_2}$&1&1&1&1&1&0&0&$6$&4&$2$ \\ \hline
  \end{tabular}}
  \end{center}\label{C2}
\end{table}
Note that each $SU(2)_{1,2}$ gauge theory is in the s-confining phase, then we find the confined description whose matter contents are anti-symmetric matters $A_{1,2}$ and singlets $Y_{SU(2)_{1,2}}$ with the $SU(4)$ gauge symmetry and the following superpotentials.
\begin{align}
W=-Y_{SU(2)_1}A_1^2 -Y_{SU(2)_2}A_2^2
\end{align}
where $A_1:= xx$ and $A_2:=\bar{x} \bar{x}$. These two singlets of the Coulomb branches couple to the singlets $S_1, S_2$ and become massive, then at low-energy we recover the electric theory.  The electric theory has the global $SU(2)_A$ symmetry which rotates the anti-symmetric matters, however in this deconfined theory this symmetry is not manifest. Only the $U(1)$ subgroup of the $SU(2)_A$ symmetry is manifest, which we denoted as $U(1)'$.

The $SU(4)$ part of the deconfined description is also s-confining, so we can move on to the s-confined description which is a 3d $\mathcal{N}=2$ $SU(2) \times SU(2)$ gauge theory:
 \begin{table}[H]\caption{Quantum numbers of the s-confined description 1} 
\begin{center}
  \begin{tabular}{|c|c|c||c|c|c|c|c|c|c| } \hline
  &$SU(2)_1$&$SU(2)_2$&$SU(2)_L$&$SU(2)_R$&$U(1)$&$U(1)$&$U(1)$&$U(1)'$&$U(1)_R$\\ \hline
$M:=Q\bar{Q}$&1&1&${\tiny \yng(1)}$&${\tiny \yng(1)}$&1&-1&4&0&$\frac{2}{3}$ \\
$M_1:=Q\bar{x}$&1&${\tiny \yng(1)}$&${\tiny \yng(1)}$&1&1&0&$\frac{1}{2}$&$-1$&$\frac{1}{3}$  \\
$M_2:=x \bar{Q}$&${\tiny \yng(1)}$&1&1&${\tiny \yng(1)}$&0&-1&$\frac{1}{2}$&1& $\frac{1}{3}$ \\
$M_3:=x \bar{x}$&${\tiny \yng(1)}$&${\tiny \yng(1)}$&1&1&0&0&-3&0&0  \\ 
$B:=x^2 Q^2$&1&1&1&1&2&0&1&2&$\frac{2}{3}$ \\
$\bar{B} :=\bar{x}^2\bar{Q}^2$&1&1&1&1&$0$&-2&1&-2&$\frac{2}{3}$ \\
$Y_{SU(4)}$&1&1&1&1&-2&2&-2&0& $ \frac{2}{3}$ \\ 
$S_1$&1&1&1&1&0&0&$-6$&$4$&$0$ \\
$S_2$&1&1&1&1&0&0&$-6$&$-4$&0  \\ \hline
$Y_{SU(2)_1}^{\mathrm{mag}}$&1&1&1&1&0&2&5&-2&$\frac{4}{3}$ \\
$Y_{SU(2)_2}^{\mathrm{mag}}$&1&1&1&1&-2&0&5&2&$\frac{4}{3}$ \\ \hline
  \end{tabular}
  \end{center}\label{qce1}
\end{table}
The superpotential is 
\begin{align}
W=S_1 Y_{SU(2)_1} +S_2  Y_{SU(2)_2}  -Y_{SU(4)} (M^2 M_3^2 -M_1^2M_2^2 +MM_1M_2M_3 -B \bar{B}).
\end{align}
The relation between the Coulomb branch operators in the electric and s-confined theories is expected from the global symmetries as
\begin{align}
Y_{SU(2)_1} &=Y_{SU(2)_1}^{\mathrm{mag}}\bar{B} \\
Y_{SU(2)_2} &= Y_{SU(2)_2}^{\mathrm{mag}} B. \label{CBrelation}
\end{align}

Both parts of the $SU(2)_{1,2}$ gauge theories are again s-confining, then we have the following s-confining description (Table \ref{SC2}).

 \begin{table}[H]\caption{Quantum numbers of the s-confined description 2} 
\begin{center}
  \begin{tabular}{|c|c||c|c|c|c|c|c|c| } \hline
  &$SU(2)_1$&$SU(2)_L$&$SU(2)_R$&$U(1)$&$U(1)$&$U(1)$&$U(1)'$&$U(1)_R$\\ \hline
$M$&1&${\tiny \yng(1)}$&${\tiny \yng(1)}$&1&-1&4&0&$\frac{2}{3}$ \\
$V:=M_1 M_1$&1&1&1&2&0&1&-2&$\frac{2}{3}$ \\
$V_1:=M_1M_3$&${\tiny \yng(1)}$&${\tiny \yng(1)}$&1&1&0&$-\frac{5}{2}$&-1&$\frac{1}{3}$ \\
$V_2:=M_3 M_3$&1&1&1&0&0&-6&0&0 \\
$M_2$&${\tiny \yng(1)}$&1&${\tiny \yng(1)}$&0&-1&$\frac{1}{2}$&1& $\frac{1}{3}$ \\
$B$&1&1&1&2&0&1&2&$\frac{2}{3}$ \\
$\bar{B} $&1&1&1&$0$&-2&1&-2&$\frac{2}{3}$   \\
$Y_{SU(4)}$&1&1&1&-2&2&-2&0& $ \frac{2}{3}$ \\
$Y_{SU(2)_2}^{\mathrm{mag}}$&1&1&1&-2&0&5&2&$\frac{4}{3}$ \\ 
$S_1$&1&1&1&0&0&$-6$&$4$&$0$ \\
$S_2$&1&1&1&0&0&$-6$&$-4$&0  \\ \hline
${Y'}_{SU(2)_1}^{\mathrm{mag}}$&1&1&1&-2&2&4&0&$\frac{2}{3}$ \\ \hline
  \end{tabular}
  \end{center}\label{SC2}
\end{table}
The superpotential describing the s-confining phase is 
\begin{align}
W&= S_1 Y_{SU(2)_1} +S_2  Y_{SU(2)_2}  -Y_{SU(4)} (M^2 V_2 -V M_2^2 +MV_1 M_2 -B \bar{B}) \nonumber \\
& \qquad \qquad -Y_{SU(2)_2}^{\mathrm{mag}} (VV_2 -V_1^2).
\end{align}
The relation of the Coulomb branch operators \eqref{CBrelation} is verified from this superpotential because if we take the equation of motion of $Y_{SU(2)_2}^{\mathrm{mag}}$, we have $S_2 B =VV_2 -V_1^2$ 
which says $S_2 x^2 Q^2 = \bar{x}^4 x^2 Q^2$. Thus $S_2$ is identified with the square of the anti-symmetric matters. The relation between the Coulomb branch operators of the $SU(2)_1$ gauge group in the two s-confined descriptions is found from the symmetry argument
\begin{align}
Y_{SU(2)_1}^{\mathrm{mag}} ={Y'}_{SU(2)_1}^{\mathrm{mag}}V.
\end{align}

The resulting $SU(2)_1$ gauge theory is also s-confining, so the theory ends up with a non-gauge theory.
The matter contents and the operator matching are summarized as follows.
 \begin{table}[H]\caption{Quantum numbers of the s-confined description 2} 
\begin{center}\scalebox{0.97}{
  \begin{tabular}{|c||c|c|c|c|c|c|c|c| } \hline
  &$SU(2)_L$&$SU(2)_R$&$U(1)$&$U(1)$&$U(1)$&$U(1)'$&$U(1)_R$ &\mbox{operator matching}\\ \hline
$M$&${\tiny \yng(1)}$&${\tiny \yng(1)}$&1&-1&4&0&$\frac{2}{3}$ & $Q\bar{Q}$\\
 $V$ &1&1&2&0&1&-2&$\frac{2}{3}$ & $AQ^2$\\
 $N:=V_1^2 $& 1&1&2&0&-5&-2 & $\frac{2}{3}$ & $\mbox{massive}$\\
$N_1:= V_1M_2$ &${\tiny \yng(1)}$&${\tiny \yng(1)}$&1&-1&-2&0&$\frac{2}{3}$& $QA^2 \bar{Q}$ \\
$N_2:=M_2M_2$&1&1&0&-2&1&2&$\frac{2}{3}$ & $A\bar{Q}^2$\\
$V_2$&1&1&0&0&-6&0&0 & $A^2$\\
$B$&1&1&2&0&1&2&$\frac{2}{3}$ & $AQ^2$\\
$\bar{B} $&1&1&$0$&-2&1&-2&$\frac{2}{3}$&$A\bar{Q}^2$ \\
$Y_{SU(4)}$&1&1&-2&2&-2&0& $ \frac{2}{3}$ &Coulomb branch\\
$Y_{SU(2)_2}^{\mathrm{mag}}$&1&1&-2&0&5&2&$\frac{4}{3}$&$\mbox{massive}$ \\
${Y'}_{SU(2)_1}^{\mathrm{mag}}$&1&1&-2&2&4&0&$\frac{2}{3}$ & Coulomb branch \\
$S_1$&1&1&0&0&$-6$&$4$&$0$ &$A^2$\\
$S_2$&1&1&0&0&$-6$&$-4$&0 & $A^2$ \\ \hline
  \end{tabular}}
  \end{center}\label{qce1}
\end{table}

The superpotential becomes
\begin{align}
W &= S_1  {Y'}_{SU(2)_1}^{\mathrm{mag}}\bar{B} V  +S_2 Y_{SU(2)_2}^{\mathrm{mag}} B  -Y_{SU(4)} (M^2 V_2 -V N_2 +MN_1 -B \bar{B}) \nonumber \\
& \qquad \qquad -Y_{SU(2)_2}^{\mathrm{mag}} (VV_2 -N) -{Y'}_{SU(2)_1}^{\mathrm{mag}}(N N_2 - N_1^2) \nonumber \\
&=   -Y_{SU(4)} (M^2 V_2 -V N_2 +MN_1 -B \bar{B}) +{Y'}_{SU(2)_1}^{\mathrm{mag}}(S_1\bar{B} V +S_2 BN_2-VV_2N_2 -N_1^2)
\end{align}
and the relation of the Coulomb branch operators is
\begin{align}
Y_{SU(2)_1} &=Y_{SU(2)_1}^{\mathrm{mag}}\bar{B} =  {Y'}_{SU(2)_1}^{\mathrm{mag}}\bar{B} V \\
Y_{SU(2)_2} &= Y_{SU(2)_2}^{\mathrm{mag}} B.
\end{align}
$N$ and $Y_{SU(2)_2}^{\mathrm{mag}}$ are massive and integrated out. The fields listed above are completely the same as \cite{Csaki:2014cwa}. However the superpotential is slightly different. Although we expect the two superpotentials describe the same physics, but it is not clear why we obtain the different potential. One of the reasons comes from the fact that our formulation does not have the explicit $SU(2)_A$ global symmetry.

This step of deconfining the anti-symmetric tensors is easily generalized to the $SU(2N)$ gauge theories with ${\tiny \yng(1,1)}$ and ${\tiny \bar{\yng(1,1)}}$. The deconfinement of the two anti-symmetric matters can be performed by introducing the $Sp(2N-2) \times Sp(2N-2)$ gauge theory. The matter contents are in Table \ref{DEC2AN}.

 \begin{table}[H]\caption{Quantum numbers of the deconfined theory of the $SU(2N)$ with anti-symmetrics} 
\begin{center}\scalebox{1}{
  \begin{tabular}{|c|c|c|c||c|c|} \hline
&$SU(2N)$&$Sp(2N-2)_1$&$Sp(2N-2)_2$&$SU(2)$&$SU(2)$ \\ \hline
$x$&${\tiny \yng(1)}$&${\tiny \yng(1)}$&1&1&1 \\
$\bar{x}$&${\tiny \bar{\yng(1)}}$&1&${\tiny \yng(1)}$&1&1 \\
$Q$&${\tiny \yng(1)}$&1&1&${\tiny \yng(1)}$&1 \\
$\bar{Q}$&${\tiny \bar{\yng(1)}}$&1&1&1&${\tiny \yng(1)}$ \\ \hline
  \end{tabular}}
  \end{center}\label{DEC2AN}
\end{table}
We omitted the global $U(1)$ charges for simplicity of the discussion and did not listed the Coulomb branch operators in the above table. The part of the $SU(2N)$ gauge theory in the deconfined description is s-confining so we have the following theory, where we will neglect the gauge singlets throughout the rest of this section.

 \begin{table}[H]\caption{Quantum numbers of the s-confined description 1} 
\begin{center}\scalebox{1}{
  \begin{tabular}{|c|c|c||c|c|} \hline
&$Sp(2N-2)_1$&$Sp(2N-2)_2$&$SU(2)$&$SU(2)$ \\ \hline
$M_1:=Q \bar{x}$&1&${\tiny \yng(1)}$&${\tiny \yng(1)}$&1 \\
$M_2:=x \bar{Q}$&${\tiny \yng(1)}$&1&1&${\tiny \yng(1)}$ \\
$M_3:=x \bar{x}$&${\tiny \yng(1)}$&${\tiny \yng(1)}$&1&1 \\ \hline
  \end{tabular}}
  \end{center}\label{qce1}
\end{table}
Again, since these two $Sp(2N-2)$ gauge theories are s-confining, the following confining description is obtained:
 \begin{table}[H]\caption{Quantum numbers of the s-confined description 2} 
\begin{center}\scalebox{1}{
  \begin{tabular}{|c|c||c|c|} \hline
&$Sp(2N-2)_1$&$SU(2)$&$SU(2)$ \\ \hline
$M_2$&${\tiny \yng(1)}$&1&${\tiny \yng(1)}$ \\
  $V_1:=M_1 M_3$&${\tiny \yng(1)}$&${\tiny \yng(1)}$&1 \\
  $V_2:=M_3M_3$&${\tiny \yng(1,1)}$&1&1 \\ \hline
  \end{tabular}}
  \end{center}\label{qce1}
\end{table}

$Sp(2N)$ gauge theory with one anti-symmetric and 4 fundamentals is special in the sense that it can be deconfined by the $Sp(2N-2)$ gauge theory and the deconfined description is again s-confining and because its s-confining description again has structure with a $Sp$ gauge theory with one anti-symmetric and 4 fundamentals. In this way we can iteratively reduce the gauge group and the ant-symmetric matters which appear in the s-confing theory can be deconfined until we end up with the $Sp(2) = SU(2)$ gauge theory which is also s-confining.  We eventually need the $N~Sp$ gauge groups to completely deconfine the theory. In this process, $N$ Coulomb branches of the $Sp$ gauge theories are introduced and these are identified with the complicated Coulomb branches of the electric $SU(2N)$ gauge theory with $2 ({\tiny \yng(1)}+\bar{{\tiny \yng(1)}}) $ and ${\tiny \yng(1,1)}+{\tiny \bar{\yng(1,1)}}$, which is consistent with the analysis in \cite{Csaki:2014cwa} and we find the 3d $\mathcal{N}=2$ $SU(2N)$ gauge theory with with ${\tiny \yng(1,1)}+{\tiny \bar{\yng(1,1)}}$ and $2~ {\tiny \yng(1,1)}$ is indeed in s-confining phase. 
Notice that for the deconfinement of the anti-symmetrics in the $Sp$ gauge group, we need to subtract the ``trace'' part of the anti-symmetric matter but this does not spoil the above discussion.

For a 3d $\mathcal{N}=2$ $SU(2N+1)$ gauge theory with $2 ({\tiny \yng(1)}+\bar{{\tiny \yng(1)}}) $, ${\tiny \yng(1,1)}$ and ${\tiny \bar{\yng(1,1)}}$, we have the following deconfined description:

\begin{table}[H]\caption{Quantum numbers of the deconfinement for $SU(2N+1)$} 
\begin{center}\scalebox{0.96}{
  \begin{tabular}{|c|c|c|c| } \hline
  &$SU(2N+1)$&$Sp(2N)_1$&$Sp(2N)_2$ \\ \hline
 $x$&${\tiny \yng(1)}$&${\tiny \yng(1)}$&1  \\
 $\bar{x}$&${\tiny \bar{\yng(1)}}$&1&${\tiny \yng(1)}$ \\
 $p_1$&1&${\tiny \yng(1)}$&1 \\
 $p_2$ &1&1&${\tiny \yng(1)}$ \\
 $Q$ &${\tiny \yng(1)}$&1&1 \\
 $\bar{Q}$&${\tiny \bar{\yng(1)}}$&1&1 \\ \hline
  \end{tabular}}
  \end{center}\label{qce1}
\end{table}
The Coulomb branches of the two $Sp(2N)$'s are lifted by introducing the singlets as before. This theory again reduces to the $Sp(2N)$ gauge theory with one anti-symmetric and 4 fundamentals and with some singlets. So we find that the theory is completely s-confining and the number of the Coulomb branch coordinates is $N$.

\section{Gauge theories with $SU(2)^N$ gauge group}
In this section, we will consider theories with product gauge groups, mainly focusing on the product of the $SU(2)$ gauge symmetries. In four dimension, these theories were studied for various reasons. One of the motivation was the $\mathcal{N}=1$ Seiberg-Witten curve \cite{Intriligator:1994sm,Csaki:1997zg,Csaki:1998dp,Gremm:1997sz} and the other was a derivation of the exact superpotential \cite{Intriligator:1994jr} by using the holomorphy and the ``linearity principle'' (integrating-in method). Here we will study the 3d version and we derive the 3d exact superpotentials and find the close relation between the Seiberg-Witten curve and the exact result superpotentials. 

The relation between the 4d Seiberg-Witten theory and the 3d $\mathcal{N}=4$ theory is investigated in \cite{Seiberg:1994aj}, in which the dimensional reduction of the curve leads to the moduli space of the 3d theory. An important claim is that the variables $x,y$ describing the Seiberg-Witten curve become physical ones if we compactify the theory on $\mathbb{R}^3 \times \mathbb{S}^1$. Here we will argue that the $\mathcal{N}=1$ Seiberg-Witten curves are also dimensionally reduced and they describe the 3d moduli space.

\subsection{Bi-fundamental and fundamental matters}
As an illustrative example, we first consider the 3d $\mathcal{N}=2$ $SU(2) \times SU(2)$ gauge theory with bi-fundamental matters and fundamental matters. The matter contents are summarize in Table \ref{PG1}. The four dimensional version of this theory was first investigated in \cite{Intriligator:1994jr}, where the authors derived an exact superpotential using the holomorphy and the integrating-in method (``linearity principle''). We here analyze the 3d version and the theory in $\mathbb{S}^1 \times \mathbb{R}^3$ and find the close relation to the 4d results. In the table below, we list the dynamical scales of the 4d theory which is necessary in $\mathbb{S}^1 \times \mathbb{R}^3$ space-time.   

\begin{table}[H]\caption{Quantum numbers of the $SU(2) \times SU(2)$ theory} 
\begin{center}\scalebox{1}{
  \begin{tabular}{|c|c|c||c|c|c|c| } \hline 
&$SU(2)_1$&$SU(2)_2$&$SU(2)$&$U(1)$&$U(1)$&$U(1)_R$  \\ \hline
$Q$&${\tiny \yng(1)}$&${\tiny \yng(1)}$&1&1&0&0  \\
$L$&1&${\tiny \yng(1)}$&${\tiny \yng(1)}$&0&1&0  \\ \hline
$Y_{SU(2)_1}$&1&1&1&$-2$&0&0  \\
$Y_{SU(2)_2}$&1&1&1&$-2$&$-2$&2  \\
$\eta_1=\Lambda_1^5$&1&1&1&2&0&2  \\
$\eta_2=\Lambda_2^4$&1&1&1&2&2&0  \\  \hline
$X:=QQ$&1&1&1&2&0&0  \\ 
$Y:=LL$&1&1&1&0&2&0  \\   \hline
   \end{tabular}}
  \end{center}\label{PG1}
\end{table}

If we assume $g_1^2 \ll g_2^2$ at some energy scale, we can replace the $SU(2)_2$ gauge theory with two flavors to the s-confined description with superpotentials
\begin{align}
W_{\mathsf{eff}}=-Y_{SU(2)_2}(XY-N^2), 
\end{align}
where we defined $N:= QL$. The low-energy effective theory is a $SU(2)_1$ gauge theory with single flavor (two doublets). The global charges are as follows.

\begin{table}[H]\caption{Quantum numbers of the $SU(2)_1$ s-confined theory} 
\begin{center}\scalebox{1}{
  \begin{tabular}{|c|c||c|c|c|c| } \hline 
&$SU(2)_1$&$SU(2)$&$U(1)$&$U(1)$&$U(1)_R$  \\ \hline
$X$&1&1&2&0&0 \\ 
$Y$&1&1&0&2&0  \\
$N:=QL$&${\tiny \yng(1)}$&${\tiny \yng(1)}$&1&1&0  \\
$Y_{SU(2)_2}$&1&1&$-2$&$-2$&2  \\  \hline
$Y'_{SU(2)_1}$&1&1&$-2$&$-2$&0  \\ \hline
   \end{tabular}}
  \end{center}\label{qcml1}
\end{table}

The Coulomb branch operator $Y'_{SU(2)_1}$ is not identical to $Y_{SU(2)_1}$ but we can easily identified as
\begin{align}
Y_{SU(2)_1} \sim Y'_{SU(2)_1} Y,
\end{align}
where the normalization factor is not important since we can absorb it by rescaling the chiral superfields.
Since the low-energy theory is $SU(2)_1$ with one flavor, it has the quantum modified moduli space \cite{Aharony:1997bx}. So, the theory is described by the following superpotential
\begin{align}
W&=-Y_{SU(2)_2}(XY-T) +\lambda (T Y'_{SU(2)_1}-1)  \label{scon1} \\
&=\lambda (XY Y'_{SU(2)_1}-1) 
\end{align}
where $T:=N^2$ and $\lambda$ is a Lagrange multiplier field. In the second line above, we integrated out $T$ and $Y_{SU(2)_2}$. It is important to note that from this analysis we first have the flat direction labeled by $N$ but finally this direction is lifted. This is very consistent with the four dimensional analysis \cite{Intriligator:1994jr}.

In order to connect the 3d dynamics to the 4d one, we need to put a theory on $\mathbb{S}^1 \times \mathbb{R}^3$ and include the non-perturbative superpotential coming from the compactification. 
This superpotential is generated by the Kaluza-Klein monopole \cite{Affleck:1982as,Aharony:1997bx,deBoer:1997kr} and the form of the superpotential takes a simple form
\begin{align}
W=\eta_1 Y_{SU(2)_1} +\eta_2  Y_{SU(2)_2}.
\end{align}
 Combining this with the s-confining superpotential \eqref{scon1}, integrating out the massive fields $T,Y_2$ and $Y'_{SU(2)_1}$ and eliminating the Lagrange multiplier $\lambda$, we arrive at 
\begin{align}
W_{\mathsf{low-energy}}=\frac{\eta_1 Y}{XY -\eta_2}.
\end{align}
This is precisely the same as \cite{Intriligator:1994jr} and we recover the 4d dynamics.

\subsection{Bi-fundamental matters}
Next we will consider the 3d $\mathcal{N}=2$ $SU(2) \times SU(2)$ gauge theory with only the bi-fundamental matters. For a single bi-fundamental, we can use the previous result by turning on the mass term to the fundamental matters. The matter contents and their quantum charges are summarized in the following table. 

\begin{table}[H]\caption{Quantum numbers} 
\begin{center}\scalebox{1}{
  \begin{tabular}{|c|c|c||c|c|c| } \hline 
&$SU(2)_1$&$SU(2)_2$&$SU(F)$&$U(1)$&$U(1)_R$  \\ \hline
$Q$&${\tiny \yng(1)}$&${\tiny \yng(1)}$&${\tiny \yng(1)}$&1&0  \\ \hline
$Y_1$&1&1&1&$-2F$&$2F-2$  \\ 
$Y_2$&1&1&1&$-2F$&$2F-2$  \\ \hline  
$\eta_1:=\Lambda_1^{6-F}$&1&1&1&$2F$& $4-2F$  \\
$\eta_2:=\Lambda_2^{6-F}$&1&1&1&$2F$& $4-2F$   \\ \hline
   \end{tabular}}
  \end{center}\label{qcml1}
\end{table}

The four dimensional version of this theory with $F=2$ is analyzed in \cite{Intriligator:1994sm} (For the generalization of the product gauge group, see \cite{Csaki:1997zg,Gremm:1997sz}). An important result of the four dimensional analysis is that the theory shows the Coulomb phase with rank 1 and can be effectively described by the $U(1)$ gauge theory whose gauge coupling is expressed by a $\mathcal{N}=1$ Seiberg-Witten curve. 

Here we will investigate the cases with $F=1$ and $F=2$ respectively and we will find the result for $F=2$ has intimate connection with the Seiberg-Witten curve obtained in \cite{Intriligator:1994sm}.
So this is showing the $\mathcal{N}=1$ generalization of \cite{Seiberg:1996nz}, where the 3d reduction of the Seiberg-Witten curve in a 4d $\mathcal{N}=2$ pure SYM with the inclusion of the softly supersymmetry breaking term.

\subsubsection{$F=1$}
Suppose the $SU(2)_2$ gauge group first becomes strong, the theory has quantum deformed moduli space 
\begin{align}
W=\lambda (XY_2-1).
\end{align}
The resulting low-energy theory has the following matter contents. $Y'_1$ is a Coulomb branch operator in the s-confined $SU(2)_1$ gauge theory.

\begin{table}[H]\caption{Quantum numbers} 
\begin{center}\scalebox{1}{
  \begin{tabular}{|c|c||c|c| } \hline 
&$SU(2)_1$&$U(1)$&$U(1)_R$  \\ \hline
$X:=QQ$&1&2&0  \\ 
$Y_2$&1&$-2$&0  \\ \hline  
$Y'_1$&1&0&$-2$  \\ \hline
   \end{tabular}}
  \end{center}\label{qcml1}
\end{table}

The $SU(2)_1$ gauge theory has no flavor and then has no stable vacua \cite{Aharony:1997bx}:
\begin{align}
W=\lambda (XY_2-1)+\frac{1}{Y'_1}.
\end{align}

This is consistent with the known result \cite{Aharony:2011ci} since the $SU(2) \times SU(2) \cong SO(4)$ and the bi-fundamental in $SU(2) \times SU(2)$ is a vector matter in $SO(4)$. In \cite{Aharony:2011ci}, it is shown that the 3d $\mathcal{N}=2$ $SO(4)$ with single flavor has no stable SUSY vacua.

For $F=1$, we can flow from the previous subsection by the mass deformation
\begin{align}
W= m LL =mY.
\end{align}
Thus the effective superpotential is 
\begin{align}
W&=-Y_{SU(2)_2}(XY-T) +\lambda (T Y'_{SU(2)_1}-1) +mY 
\end{align}
and this has no stable solution.

\subsubsection{$F=2$}
For $F=2$, each dynamics of $SU(2)_{1,2}$ factors is in the s-confining phase. First we will change the $SU(2)_2$ dynamics. Since the low-energy theory (see Table \ref{BI2}) involves only the adjoint matter, we have the enhanced $\mathcal{N}=4$ supersymmetry if we omit the superpotential
\begin{align}
W= -Y_2(X_1^2 +X_2^2), 
\end{align}
which is breaking the supersymmetry from $\mathcal{N}=4$ to $\mathcal{N}=2$.

\begin{table}[H]\caption{Quantum numbers of the low-energy $SU(2)_1$ theory} 
\begin{center}\scalebox{1}{
  \begin{tabular}{|c|c||c|c|c| } \hline 
&$SU(2)_1$&$SU(F=2)$&$U(1)$&$U(1)_R$  \\ \hline
$X_1:=QQ$&adj.&1&2&0  \\
$X_2:=QQ$&1&adj.&2&0 \\
$Y_2$&1&1&$-4$&2  \\ \hline  
$Y'_1$&1&1&$-4$&0  \\ \hline
   \end{tabular}}
  \end{center}\label{BI2}
\end{table}

The resulting theory is a $SU(2)$ gauge theory with an adjoint matter, which is regarded as the $SO(3)$ gauge theory with one fundamental (vector) matter. This theory is known to be described by the following quantum constraint \cite{Aharony:2011ci}
\begin{align}
Y^{2}_{SO(3)} N +\tilde{q}^2 +1=0,
\end{align}
where $N:=X_1^2$ and $Y'_1=Y^{2}_{SO(3)}$. The relation of the Coulomb branch operators between the $SU(2)$ and the $SO(3)$ gauge groups is very plausible since the minimal monopole charges are different in these gauge groups. Furthermore the global charges of these two operators suggest this relation. Since we are actually dealing with not the $SO(3)$ but the $SU(2)$ gauge group, it would be plausible to change the variables
\begin{align}
Y_{SO(3)}^2 \rightarrow Y'_{1},~~~~~Y_{SO(3)} \tilde{q} \rightarrow q.
\end{align}
Under this change, the constraint becomes 
\begin{align} 
Y'^2_1 N+q^2 +Y'_1 =0.
\end{align}
Summing up all the superpotentials and the constraint, we have the low-energy dynamics described by
\begin{align}
W &=-Y_2(N+X_2^2) +\lambda(Y'^2_1 N+q^2 +Y'_1 ) \nonumber \\
& \rightarrow \lambda(-Y'^2_1 X_2^2 +\tilde{q}^2+Y'_1),
\end{align}
where in the second line we have integrated out the massive fields $Y_2$ and $N$.
This is again consistent with the analysis of \cite{Aharony:2011ci}, where a 3d $\mathcal{N}=2$ $O(4)$ theory with two flavors is investigated and they claim that the Coulomb phase is described by the similar constraint.

Let us connect the 3d dynamics with the 4d one. The 4d dynamics was studied in \cite{Intriligator:1994sm}. The low-energy theory is in the Coulomb phase with rank $1$ and the gauge coupling and the singularity structure on the moduli space are captured by the following Seiberg-Witten curve
\begin{align}
y^2 =x^3 +x^2 (-U+\Lambda_1^4 +\Lambda_2^4) +\Lambda_1^4\Lambda_2^4 x,
\end{align}
where $U :=\det QQ$. In order to dimensionally reduce the theory to 3d, we first notice that the $x,y$ variables are not physical ones in the 4d limit but in $\mathbb{R}^3 \times \mathbb{S}^1$ or in the 3d limit we can treat these variables as physical fields \cite{Seiberg:1996nz}. Then we can introduce the Seiberg-Witten curve as the constraints between the chiral superfields $x,y$ and $U$. Notice that $U$ is a physical variable in 4d and 3d, along which we have the massless excitation.

When we take the 3d limit we should redefine the variables $x, y, U$ and absorb $\Lambda_1$ and $\Lambda_2$ since in the 3d limit we must take $\Lambda_{1,2} \rightarrow 0$ and the three dimensional gauge couplings to be fixed. According to the Seiberg-Witten prescription \cite{Seiberg:1996nz} we change the variables as
\begin{align}
\tilde{x} &:=(\Lambda_1^4\Lambda_2^4)^{-1}x  \\
\tilde{y}  &:=(\Lambda_1^4\Lambda_2^4)^{-1} y \\
v&:=x-U+\Lambda_1^4+\Lambda_2^4,
\end{align}
then the Seiberg-Witten curve becomes 
\begin{align}
\tilde{y}^2 =\tilde{x}^2 v +\tilde{x}.
\end{align}
In this expression we can safely take the 3d limit and find the good agreement with the 3d analysis with the identification $\tilde{y}=\tilde{q}$, $v=X_2^2$ and $\tilde{x}=-Y'_1$.

\subsection{$SU(2)\times SU(2) \times SU(2)$}
We generalize the previous result on the $SU(2) \times SU(2)$ with bi-fundamental matters to the case with  a $SU(2) \times SU(2) \times SU(2)$ gauge symmetry. The four dimensional version is investigated in \cite{Csaki:1997zg}, where the Seiberg-Witten curve is derived. The matter contents and their global charges are in Table \ref{triple}.

\begin{table}[H]\caption{Quantum numbers of the triple $SU(2)$'s theory} 
\begin{center}\scalebox{1}{
  \begin{tabular}{|c|c|c|c||c|c|c|c| } \hline 
&$SU(2)_1$&$SU(2)_2$&$SU(2)_3$&$U(1)_1$&$U(1)_2$&$U(1)_3$&$U(1)_R$ \\ \hline
$Q_1$&${\tiny \yng(1)}$&${\tiny \yng(1)}$&1&1&0&0&0 \\
$Q_2$&1&${\tiny \yng(1)}$&${\tiny \yng(1)}$&0&1&0&0  \\
$Q_3$&${\tiny \yng(1)}$&1&${\tiny \yng(1)}$&0&0&1&0 \\ \hline
$Y_1$&1&1&1&$-2$&0&$-2$&2  \\
$Y_2$&1&1&1&$-2$&$-2$&0&2  \\
$Y_3$&1&1&1&$0$&$-2$&$-2$&2  \\  \hline
   \end{tabular}}
  \end{center}\label{triple}
\end{table}
Since all the gauge groups have the structure with the $SU(2)$ with two flavors, the theory is reduced to the $SU(2) \times SU(2)$ theory by going to the s-confined description. If we first change the $SU(2)_3$, we obtain the following theory in Table \ref{triple1}, where we put primes on the Coulomb branch coordinates to distinguish these operators from $Y_{1,2}$.

\begin{table}[H]\caption{Quantum numbers of the $SU(2)_1 \times SU(2)_2$ theory} 
\begin{center}\scalebox{1}{
  \begin{tabular}{|c|c|c||c|c|c|c| } \hline 
&$SU(2)_1$&$SU(2)_2$&$U(1)_1$&$U(1)_2$&$U(1)_3$&$U(1)_R$ \\ \hline
$Q_1$&${\tiny \yng(1)}$&${\tiny \yng(1)}$&1&0&0&0 \\
$N_1:=Q_2Q_3$&${\tiny \yng(1)}$&${\tiny \yng(1)}$&0&1&1&0  \\
$M_2:=Q_2Q_2$&1&1&0&2&0&0  \\
$M_3:=Q_3Q_3$&1&1&0&0&2&0  \\
$Y_3$&1&1&0&$-2$&$-2$&2  \\  \hline 
$Y'_1$&1&1&$-2$&$-2$&$-2$&2  \\
$Y'_2$&1&1&$-2$&$-2$&$-2$&2  \\  \hline
   \end{tabular}}
  \end{center}\label{triple1}
\end{table}
The superpotential is
\begin{align}
W=-Y_3(N_1^2+M_2M_3)
\end{align}
and the relations between the Coulomb branch operators are
\begin{align}
Y'_1 M_2  = Y_1,~~~Y'_2 M_3=Y_2.
\end{align}

The matter contents charged under the $SU(2)_1 \times SU(2)_2$ are completely the same as the previous subsection so we can reduce the $SU(2) \times SU(2)$ gauge symmetry to a single $SU(2)$. The low-energy superpotential is
\begin{align}
W= -Y'_2 (M_1 N_2+N_3^2+N_4^2).
\end{align}

\begin{table}[H]\caption{Quantum numbers of the $SU(2)_1$ theory} 
\begin{center}\scalebox{1}{
  \begin{tabular}{|c|c||c|c|c|c| } \hline 
&$SU(2)_1$&$U(1)_1$&$U(1)_2$&$U(1)_3$&$U(1)_R$ \\ \hline
$M_1:=Q_1Q_1$&1&2&0&0&0  \\
$N_2:=N_1N_1$&1&0&2&2&0  \\
$N_3:=Q_1N_1|_{\mathrm{anti-symmetric\,part}} $&1&1&1&1&0  \\
$N_4:=Q_1N_1|_{\mathrm{symmetric\,part}} $&$\mathrm{adj.}$&1&1&1&0  \\
$Y_3$&1&0&$-2$&$-2$&2  \\
$Y'_2$&1&$-2$&$-2$&$-2$&2  \\  \hline
$Y''_1$&1&$-2$&$-2$&$-2$& 0   \\ \hline
   \end{tabular}}
  \end{center}\label{qcml1}
\end{table}
Since this theory is again the same as the previous subsection, so we have the quantum modified constraint as the low-energy description
\begin{align}
Y''^2_{1} T +\tilde{q}^2 +Y''_1 =0,  
\end{align}
where $T:=N_4^2$. Combining all the superpotentials we end up with
\begin{align}
W&=-Y_3(N_1^2+M_2M_3)-Y'_2 (M_1 N_2+N_3^2+N_4^2) +\lambda(Y''^2_{1} T +\tilde{q}^2 +Y''_1) \nonumber \\
&=-Y_3 (N_2 +M_2M_3)-Y'_2 (M_1 N_2+N_3^2+T) +\lambda(Y''^2_{1} T +\tilde{q}^2 +Y''_1). 
\end{align}
Integrateing out $Y_3, N_2, Y'_2$ and $T$, we obtain the low-energy effective superpotential.
\begin{align}
W=\lambda(Y''^2_{1}  (M_1M_2M_3-N_3^2) +\tilde{q}^2 +Y''_1)
\end{align}
This is again consistent with the Seiberg-Witten curve describing the four dimensional version of this theory
\begin{align}
y^2 =x^3 +x^2 (\Lambda_1^4 M_2 +\Lambda_2^4 M_3 +\Lambda_3^4 N_1 -N_1M_2 M_3 +N_3^2) +x \Lambda_1^4\Lambda_2^4  \Lambda_3^4.
\end{align}

We can easily generalize the above analysis to the 3d $\mathcal{N}=2$ $SU(2)^N$ gauge theory and obtain the low-energy effective description
\begin{align}
W=\lambda (Y^2 (M_1 \cdots M_N -T^2) +\tilde{q}^2 +Y),
\end{align}
where $M_i:=Q_i Q_i$ and $T:=Q_1 \cdots Q_N$. One possible application of the 3d $SU(2)^{N}$ theory is an analysis of the 3d $\mathcal{N}=2$ $SU(4)$ with $4$ anti-symmetrics. This theory can be deconfined into the 3d $\mathcal{N}=2$ $SU(4) \times SU(2)^4$ gauge theory, in which the Coulomb branches of the $SU(2)^4$ should be lifted. The $SU(4)$ part is also in the s-confining phase and the low-energy theory reduces to the $SU(2)^4$ theory discussed in this section.

\section{$SU(N)$ with an antisymmetric matter}
In this section we will consider the deconfinement and the Seiberg dual of the theory with an anti-symmetric matter, where the deconfined description does not have the ``s-confining $\times$ s-confining'' structure. Although the anti-symmetric matter is deconfined in the same way as the previous sections by using the dynamics of the 3d $\mathcal{N}=2$ $Sp$ gauge theory, constructing the dual theory will become very complicated because in our method we are lifting up the unnecessary Coulomb moduli and the Coulomb branch operator labeling it is non-trivially mapped under the duality transformation. Then we here show two deconfinement descriptions and their duals. First, we will deconfine the anti-symmetric by lifting up the Coulomb moduli of the $Sp$ gauge group. This is completely the same as the previous one. Next we will give the different approach, where the un-wanted Coulomb branch is just decoupled from the other sector and remaining as the massless direction. 
\subsection{Lifting up the Coulomb moduli}
\subsubsection{The electric theory}
The electric theory is a 3d $\mathcal{N}=2$ $SU(2N)$ gauge theory with $F \, {\tiny \yng(1)}$, $F' \, \bar{{\tiny \yng(1)}}$ and ${\tiny \yng(1,1)}$ with no tree-level superpotential. Since there is no chiral anomaly in 3d, any values of $F$ and $F'$ are allowed. The generalization to a $SU(2N+1)$ case is straightforward, so we will not discuss it. The quantum numbers of the matter contents are as follows.

\begin{table}[h]\caption{Quantum numbers of the electric theory} 
\begin{center}
  \begin{tabular}{|c|c||c|c|c|c|c|c| } \hline
     &$SU(2N)$& $SU(F)$ & $SU(F')$ & $U(1)_B$ & $U(1)_A$ & $U(1)_X$ & $U(1)_R$  \\ \hline
   $Q$  & $\Box$  &  $\Box$ &1&1&1&0&0 \\
   $\tilde{Q}$& $\bar{{\tiny \yng(1)}}$ &  1& $\Box$ &  $-1$&1&0&0 \\
  $A$ & ${\tiny \yng(1,1) }$&1&1&0&0&2&0 \\ \hline
  \end{tabular}
  \end{center}\label{SDwae}
\end{table}
The global $U(1)_R$ charge can be generally mixed with other $U(1)$ symmetries and the genuine $U(1)_R$ charge which is realized in the IR fixed point would not be the above combination. 
In this section we will consider the dual description for $F'=F+2N-2$ in which we can use the 3d $SU(N_c)$ Seiberg duality developed in \cite{Aharony:2013dha}.

\subsubsection{The deconfined description}
The anti-symmetric matter can be deconfined into the 3d $\mathcal{N}=2$ $SU(2N) \times Sp(2N-2)$ gauge theory with the following matter contents (Table \ref{qced1}):
\begin{table}[H]\caption{Quantum numbers of the deconfined theory} 
\begin{center}\scalebox{0.9}{
  \begin{tabular}{|c|c|c||c|c|c|c|c|c| } \hline
     &$SU(2N)$&Sp(2N-2)& $SU(F)$ & $SU(F')$& $U(1)_B$ & $U(1)_A$ & $U(1)_X$ & $U(1)_R$  \\ \hline
     $Q$  & $\Box$  & $1$ & $\Box$ &1&1&1&0&0  \\
     $\tilde{Q}$& $\bar{{\tiny \yng(1)}}$ & $1$&  1& $\Box$ &  $-1$&1&0&0 \\
     $x$& $\Box$&$\Box$&1&1&0&0&$1$& 0  \\
     $S$&1&1&1&1&$0$&$0$&$2N$&$0$\\ \hline
     $Y_{SU(2N)}$&1&1&1&1&$2N-2$&$-2F-2N+2$&$-2N+2$&$2F-2$ \\
     $Y_{Sp(2N-2)}$&1&1&1&1&0&0&$-2N$&2  \\ \hline
    \end{tabular}}
  \end{center}\label{qced1}
\end{table}
The difference of the 4d and 3d cases is the appearance of the Coulomb branch $Y_{Sp(2N-2)}$ and the new singlet $S$ which couples to the unnecessary Coulomb branch direction and lifts up it. Then the deconfined theory has to include the superpotential
\begin{align}
W=SY_{Sp(2N-2)}.
\end{align}

Applying the s-confining description of the $Sp(2N-2)$ gauge theory when $g^2_{SU} \ll g^2_{Sp}$, we go back to the electric theory mentioned above. The effective low-energy degrees of freedom are the mesons $A=xx$ and the Coulomb branch coordinate $Y_{Sp(2N-2)} $. However, the field $Y_{Sp(2N-2)}$ is massive due to the superpotential.

\subsubsection{The first dual description}
For $g^2_{SU} \gg g^2_{Sp}$, we can think of the $Sp(2N-2)$ gauge group as the flavor symmetry and we can use the 3d Seiberg duality for $SU(N_c)$ gauge group \cite{Aharony:2013dha} if we restrict ourself to the case with $F'=F+2N-2$.
The dual description is given by a 3d $\mathcal{N}=2$ $U(F-2) \times Sp(2N-2)$ gauge theory with following matter contents;
\begin{table}[H]\caption{Quantum numbers of the first dual theory} 
\begin{center}\scalebox{0.8}{
  \begin{tabular}{|c|c|c||c|c|c|c|c|c| } \hline
     &$U(F-2)$&$Sp(2N-2)$& $SU(F)$ & $SU(F+2N-2)$ & $U(1)_B$ & $U(1)_A$ & $U(1)_X$ & $U(1)_R$  \\ \hline
     $q$&${\tiny \yng(1)}_1$&1&${\tiny \bar{\yng(1)}}$&1&0&$-1$&0&1 \\
     $\tilde{q}$&${\tiny \bar{\yng(1)}}_{-1}$&1&1&${\tiny \bar{\yng(1)}}$&0&$-1$&0&1 \\
     $\tilde{x}$&${\tiny \yng(1)}_{1}$&${\tiny \yng(1)}$&1&1&1&0&$-1$&1 \\
     $b$&$1_{-(F-2)}$&1&1&1&2&$F$&$2N-2$&$2-F$ \\
     $\tilde{b}$&$1_{+(F-2)}$&1&1&1&$-2N$&$F+2N-2$&0&$2-F$ \\
     $Y_{SU(2N)}$&1&1&1&1&$2N-2$&$-2F-2N+2$&$2-2N$&$2F-2$ \\
     $M:=Q\tilde{Q}$&1&1&${\tiny \yng(1)}$&${\tiny \yng(1)}$&0&2&0&0 \\
     $M_1:=x \tilde{Q}$&1&${\tiny \yng(1)}$&1&${\tiny \yng(1)}$&$-1$&1&1&0 \\   \hline
     $\tilde{X}_{\pm}$&1&1&1&1&0&0&0&2  \\ 
     $Y_{Sp(2N-2)}^{\mathrm{mag}}$&$1_{-(F-2)}$&1&1&1&$2N$&$-F-2N+2$&$-2N$&$F$  \\ \hline
  \end{tabular}}
  \end{center}\label{qcm1}
\end{table}
\noindent where $\tilde{X}_{\pm}$ are the Coulomb branch coordinates of the $U(F-2)$ gauge group and the $Y_{Sp(2N-2)}^{\mathrm{mag}}$ is describing the $Sp(2N-2)$. For the non-zero vevs of $\tilde{X}_{\pm}$ break the gauge symmetry as $U(F-2) \rightarrow U(F-4) \times U(1) \times U(1)$ and the operators $\tilde{X}_{\pm}$ create the monopoles corresponding to $U(1) \times U(1)$.
We can not identify $Y_{Sp(2N-2)}^{\mathrm{mag}}$ with the Coulomb branch operator $Y_{Sp(2N-2)}$ of the deconfined theory. 
The first dual description has the superpotential
\begin{align}
W=M q \tilde{q} +M_1 \tilde{x} \tilde{q} +Y_{SU(2N)} b \tilde{b} +\tilde{X}_+ +\tilde{X}_{-},
\end{align}
which is consistent with all the global symmetries. 


%
The matching of the baryonic operators between the electric and magnetic theory is
\begin{align}
B_l &:=x^l Q^{2N-l} =b \tilde{x}^{2N-2-l} q^{F+l-2N} \\
\bar{B} &:= \tilde{Q}^{2N} =\tilde{b} \bar{q}^{F-2}
\end{align}
and this identification removes almost all the ambiguities of the assignment of the $U(1)$ charges of the dual theory. A subtlety of the operator matching between the dual and the deconfined theories arises from the $Sp(2N-2)$ Coulomb branch operators. If we regard the $U(F-2)$ gauge symmetry as global one, the $Sp(2N-2)$ Coulomb branch operator $Y_{Sp(2N-2)}^{\mathrm{mag}}$ is negatively charged under the $U(1) \subset U(F-2)$. So we can make it neutral by multiplying some fields. In this case the $Sp(2N-2)$ Coulomb branch operators are expected to be identified as
\begin{align}
Y_{Sp(2N-2)} = Y_{Sp(2N-2)}^{\mathrm{mag}} \tilde{b},
\end{align}
which is completely consistent with other global symmetries. 

Next we can perform the Aharony dual transformation for the $Sp(2N-2)$ gauge theory \cite{Aharony:1997gp} and the second dual theory will be given by the 3d $\mathcal{N}=2$ $U(F-2) \times Sp(2F-4)$ gauge theory. However the theory includes the anti-symmetric matter in the $U(F-2)$ gauge group and its Coulomb brach becomes very complicated. We will generally expect to need two coordinates for describing it. It is highly non-trivial how to map the operators $\tilde{X}_{\pm}$. Since these are appearing in the superpotential, we must translate these operators in terms of the second dual theory language.

\subsubsection{The second dual description}
The second dual description is given by a 3d $\mathcal{N}=2$ $U(F-2) \times Sp(2F-4)$ gauge theory. The matter contents are in Table \ref{SECD} and the superpotential is
\begin{align}
W&=M q \tilde{q} +N_1 \tilde{q} +Y_{SU(2N)} b \tilde{b} +\tilde{X}_+ +\tilde{X}_{-}\nonumber \\
&\qquad +N\tilde{\tilde{x}}\tilde{\tilde{x}}+N_1\tilde{\tilde{x}}\tilde{M}_1 +N_2 \tilde{M}_1\tilde{M}_1
+Y_{Sp(2N-2)}^{\mathrm{mag}} Y_{Sp(2F-4)} +SY_{Sp(2N-2)}^{\mathrm{mag}} \tilde{b}.
\end{align}
Notice that $N_1$ and $\tilde{q}$ are massive and integrating over these fields induces the fourth order term in the superpotential.

\begin{table}[H]\caption{Quantum numbers of the second dual theory} 
\begin{center}\scalebox{0.782}{
  \begin{tabular}{|c|c|c||c|c|c|c|c|c| } \hline
     &$U(F-2)$&$Sp(2F-4)$& $SU(F)$ & $SU(F+2N-2)$ & $U(1)_B$ & $U(1)_A$ & $U(1)_X$ & $U(1)_R$  \\ \hline
     $q$&${\tiny \yng(1)}_1$&1&${\tiny \bar{\yng(1)}}$&1&0&$-1$&0&1 \\
     $\tilde{q}$&${\tiny \bar{\yng(1)}}_{-1}$&1&1&${\tiny \bar{\yng(1)}}$&0&$-1$&0&1 \\
     $\tilde{\tilde{x}}$&${\tiny \bar{\yng(1)}}_{-1}$&${\tiny \yng(1)}$&1&1&$-1$&0&1&0 \\
     $b$&$1_{-(F-2)}$&1&1&1&2&$F$&$2N-2$&$2-F$ \\
     $\tilde{b}$&$1_{+(F-2)}$&1&1&1&$-2N$&$F+2N-2$&0&$2-F$ \\
     $Y_{SU(2N)}$&1&1&1&1&$2N-2$&$-2F-2N+2$&$2-2N$&$2F-2$ \\
     $M:=Q\tilde{Q}$&1&1&${\tiny \yng(1)}$&${\tiny \yng(1)}$&0&2&0&0 \\
     $\tilde{M}_1$&1&${\tiny \yng(1)}$&1&${\tiny \bar{\yng(1)}}$&1&$-1$&$-1$&1 \\   
     $N:=\tilde{x}\tilde{x}$&${\tiny \yng(1,1)}_2$&1&1&1&2&0&$-2$&2  \\
     $N_1:=\tilde{x}M_1$&${\tiny \yng(1)}_1$&1&1&${\tiny \yng(1)}$&0&1&0&1  \\
     $N_2:=M_1M_1$&1&1&1&${\tiny \yng(1,1)}$&$-2$&2&2&0  \\ 
     $Y_{Sp(2N-2)}^{\mathrm{mag}}$&$1_{-(F-2)}$&1&1&1&$2N$&$-F-2N+2$&$-2N$&$F$  \\ \hline
     $\tilde{X}_{\pm}$&1&1&1&1&0&0&0&2  \\ 
     $Y_{Sp(2F-4)}$&$1_{+(F-2)}$&1&1&1&$-2N$&$F+2N-2$&$2N$&$-F+2$  \\  \hline
  \end{tabular}}
  \end{center}\label{SECD}
\end{table}

The problem of this description is that since the part of $U(F-2)$ now includes the anti-symmetric matter ${\tiny \yng(1,1)}_2$, the corresponding Coulomb branch becomes highly involved. In general, we expect more than one coordinate to describe the Coulomb moduli. Here we start with the analysis of the Coulomb modui from the lower values of $F$. This analysis is inevitably required because the superpotential contains the Coulomb branch operator $\tilde{X}_\pm$ and these operators are the ones of the first dual theory and we need to express these operators in the second dual theory language.

For $F=3$ where the gauge group is $U(1)$ and all the fields have the charges $\pm1$ under this $U(1)$. Notice that we cannot construct $N:=\tilde{x}\tilde{x}$ for $F=3$. This very simplifies the structure of the Coulomb branch. All the charged fermions are equally contributing to the zero-modes around the monopole background with respect to this $U(1)$. The global charges of the Coulomb branch operators $\tilde{X}_{\pm}^{U(1)}$ which is related to the above monopole can be calculated by counting these fermionic zero-modes. Thus we find that the Coulomb branch operators $\tilde{X}_{\pm}^{U(1)}$ has the same global charges as $\tilde{X}_{\pm}$. Therefore $\tilde{X}_{\pm}$ are naturally mapped to the Coulomb branch operators $\tilde{X}^{U(1)}_{\pm}$ of the second dual $U(1)\times Sp(2)$ gauge theory up to an irrelevant normalization factor.

For $F=4$, We have a $U(2)$ gauge theory with the doublets $F\, \Box_1,~(2F-4)\, \Box_{-1}$, two negatively charged singlets $1_{-2}$ and two positively charged singlets $1_{+2}$, where we are not including the massive matters. Along the Coulomb branch, the $U(2)$ gauge symmetry breaks to $U(1)_1 \times U(1)_2$ and the matter fields decompose as
\begin{align}
\Box_1 &\rightarrow 1_{(1,0)} +1_{(0,1)},~~~ \Box_{-1} \rightarrow 1_{(1,0)} +1_{(0,1)} \\
 1_{-2} & \rightarrow 1_{(-1,-1)},~~~~~~~~~ 1_{+2} \rightarrow 1_{(1,1)},
\end{align}
where we renoramalized the $U(1)$'s by reorganizing the Cartan subalgebra of $U(2)$ and these $U(1)$'s are defined as
\begin{align}
T_{1}=\left(
    \begin{array}{cc}
      1 & 0  \\
      0 & 0  \\
    \end{array}
  \right),~~~T_2  =\left(
    \begin{array}{cc}
      0 & 0  \\
      0 & 1  \\
    \end{array}
  \right).
\end{align}
The above decomposition of the matters implies that the all the matter fields are equally contributing the mixed Chern-Simons terms such as
\begin{align}
 k_{\mathrm{eff}}^{U(1)_1,U(1)_B}=\frac{1}{2} \sum_{{\tiny \mbox{all the fermions}}}  q_{U(1)_1} q_{U(1)_B} \mathrm{sign}(m) 
\end{align}
which determines the $U(1)_B$ charges of the Coulomb branch operators $\tilde{X}^{U(2)}_{\pm}$. Along the Coulomb branch with $U(2) \rightarrow U(1) \times U(1)$, we find 
\begin{align}
k_{\mathrm{eff}}^{U(1)_1,U(1)_B}=k_{\mathrm{eff}}^{U(1)_1,U(1)_A}=k_{\mathrm{eff}}^{U(1)_1,U(1)_X}=0,~~~k_{\mathrm{eff}}^{U(1)_1,U(1)_R}=-2.
\end{align}
From the relation between the mixed Chern-Simons terms and global $U(1)$ charges,
\begin{align}
q^{U(1)_{global}} =-k_{\mathrm{eff}}^{U(1)_1,U(1)_{global}},
\end{align}
we find that the quantum numbers of the Coulomb branch operator $\tilde{X}_{\pm}^{U(2)}$ are identical to the ones of $\tilde{X}_{\pm}$. We end up with the identification $\tilde{X}_{\pm} \sim \tilde{X}^{U(2)}_{\pm}$ up to normalizations.

For $F=5$, we have the $U(3)$ gauge theory with the ``chiral'' matters and some singlets under the $SU(3) \subset U(3)$. Since these matter fields have the different $U(1) \sim U(3)/SU(3)$ charges, the Coulomb branch has the complicated quantum numbers. Under the breaking with $U(3) \rightarrow U(1)_1 \times U(1)_2 \times U(1)_3$, the matter field ${\tiny \bar{\yng(1)}}_2$ decomposes into
\begin{align}
1_{(0,1,1)}+1_{(1,0,1)}+1_{(1,1,0)}.
\end{align}
This means that the effective mixed Chern-Simons term $k_{\mathrm{eff}}^{U(1)_1,U(1)_{others}}$ has two contributions from this matter field. Noticing this subtlety we again find $\tilde{X}_{\pm} \sim \tilde{X}^{U(3)}_{\pm}$.

For $F>5$, the $U(F-2)$ theory has the anti-symmetric matter ${\tiny \yng(1,1)}_2$ and the Coulomb branch becomes more complicated. Then we further deconfine this anti-symmetric matter by introducing the $Sp(F-4)$ gauge theory, where we are assuming $F$ is even, but it is easily generalized to the odd $F$ case which we will not discuss here. The deconfined theory has the following matter contents and their global charges (Table \ref{DEC2nd}).

\begin{table}[H]\caption{Quantum numbers of the second deconfined theory} 
\begin{center}\scalebox{0.7}{
  \begin{tabular}{|c|c|c|c||c|c|c|c|c|c| } \hline
     &$U(F-2)$&$Sp(2F-4)$&$Sp(F-4)$& $SU(F)$ & $SU(F+2N-2)$ & $U(1)_B$ & $U(1)_A$ & $U(1)_X$ & $U(1)_R$  \\ \hline
     $q$&${\tiny \yng(1)}_1$&1&1&${\tiny \bar{\yng(1)}}$&1&0&$-1$&0&1 \\
     $\tilde{\tilde{x}}$&${\tiny \bar{\yng(1)}}_{-1}$&${\tiny \yng(1)}$&1&1&1&$-1$&0&1&0 \\
     $b$&$1_{-(F-2)}$&1&1&1&1&2&$F$&$2N-2$&$2-F$ \\
     $\tilde{b}$&$1_{+(F-2)}$&1&1&1&1&$-2N$&$F+2N-2$&0&$2-F$ \\
     $Y_{SU(2N)}$&1&1&1&1&1&$2N-2$&$-2F-2N+2$&$2-2N$&$2F-2$ \\
     $M:=Q\tilde{Q}$&1&1&1&${\tiny \yng(1)}$&${\tiny \yng(1)}$&0&2&0&0 \\
     $\tilde{M}_1$&1&${\tiny \yng(1)}$&1&1&${\tiny \bar{\yng(1)}}$&1&$-1$&$-1$&1 \\   
     $y$&${\tiny \yng(1)}_1$&1&${\tiny \yng(1)}$&1&1&1&0&$-1$&1  \\
     $N_2:=M_1M_1$&1&1&1&1&${\tiny \yng(1,1)}$&$-2$&2&2&0  \\ 
     $Y_{Sp(2N-2)}^{\mathrm{mag}}$&$1_{-(F-2)}$&1&1&1&1&$2N$&$-F-2N+2$&$-2N$&$F$  \\ 
     $S'$&$1_{+(F-2)}$&1&1&1&1&$F-2$&$0$&$2-F$&$F-2$ \\ \hline
     $\tilde{X}_{\pm}$&1&1&1&1&1&0&0&0&2  \\ 
     $Y_{Sp(2F-4)}$&1&1&1&1&1&$-2N$&$F+2N-2$&$2N$&$-F+2$  \\  
     $Y_{Sp(F-4)}$&$1_{-(F-2)}$&1&1&1&1&$2-F$&0&$F-2$&$4-F$  \\
     $X_{\pm}^{\mathrm{dec}}$&1&1&1&1&1&0&0&0&2  \\ \hline
  \end{tabular}}
  \end{center}\label{DEC2nd}
\end{table}
The anti-symmetric matter $N$ is deconfined into $y$ and we introduce a singlet in order to lift up the Coulomb moduli of $Sp(F-4)$. The important point is that the Coulomb moduli $Y_{Sp(F-4)}$ is charged under the U(1), the abelian part of $U(F-2)$. Then $S'$ is also charged under the $U(F-2)$ and contributes as the zero-mode around the monopole background. The superpotential now includes $\delta W=S' Y_{Sp(F-4)}$ and this term is very non-trivial from the UV field theory point of view. In order to avoid this difficulty, we will discus the different approach of decanting the two-index matters in the next subsection. Summing up the all the contributions to the superpotential, we have
 
\begin{align}
W&=-M q \tilde{\tilde{x}}\tilde{M}_1  +Y_{SU(2N)} b \tilde{b} +\tilde{X}_+ +\tilde{X}_{-}\nonumber \\
&\qquad +(yy)\tilde{\tilde{x}}\tilde{\tilde{x}} +N_2 \tilde{M}_1\tilde{M}_1
+Y_{Sp(2N-2)}^{\mathrm{mag}} Y_{Sp(2F-4)} +SY_{Sp(2N-2)}^{\mathrm{mag}} \tilde{b} +S' Y_{Sp(F-4)}.
\end{align}
We can easily find that $\tilde{X}_{\pm}$ are mapped to the Coulomb branch operators $X_{\pm}^{\mathrm{dec}}$ of the second deconfined $U(F-2) \times Sp(2F-4) \times Sp(F-4)$ gauge theory.

\subsection{Decoupling the Coulomb moduli}
Alternatively we will show the different approach, where the unnecessary Coulomb branch is not lifted but remains as the flat direction and it will be completely decoupled from the other sector.
Let us again consider the 3d $\mathcal{N}=2$ $SU(N)$ gauge theory with single anti-symmetric, $F$ fundamentals and $F'=F+2N-2$ anti-fundamentals without tree-level superpotentials. This theory is deconfined into the following theory in Table \ref{DEC1} without the introduction of the singlet lifting up the Coulomb branch. We need instead include the additional matter fields and the superpotential
\begin{align}
W=xp_1 p_2 +p_1p_1p_3.
\end{align}

\begin{table}[h]\caption{Quantum numbers of the deconfined theory} 
\begin{center}\scalebox{0.87}{
  \begin{tabular}{|c|c|c||c|c|c|c|c|c|c| } \hline
     &$SU(N)$&Sp(N+K-2)& $SU(F)$ & $SU(F')$ &SU(K)& $U(1)_B$ & $U(1)_A$ & $U(1)_X$ & $U(1)_R$  \\ \hline
   $Q$  & $\Box$  & $1$ & $\Box$ &1&1&1&1&0&0 \\
   $\tilde{Q}$& $\bar{{\tiny \yng(1)}}$ & $1$&  1& $\Box$ &1&  $-1$&1&0&0 \\
$x$& $\Box$&$\Box$&1&1&1&0&0&$\frac{1}{2}$& 0 \\
$p_1$&1&$\Box$&1&1&$\Box $&$B$&$A$&$X$&$R$ \\
$p_2$ &$\bar{ {\tiny \yng(1)}}$ &1&1&1& $\bar{ {\tiny \yng(1)}}$&$-B$&$-A$&$-\frac{1}{2}-X$& $2-R$\\
$p_3$&1&1&1&1& $\bar{{\tiny \yng(1,1)}}$&$-2B$&$-2A$&$-2X$&$2-2R$ \\
$S$&1&1&1&1&1&$BK$&$AK$&$\frac{1}{2}N+XK$&$RK$ \\ \hline 
$Y_{sp}$&1&1&1&1&1&$-BK$&$-AK$&$-\frac{1}{2}N-XK$&$-RK+2$ \\ \hline
  \end{tabular}}
  \end{center}\label{DEC1}
\end{table}

Notice that in this theory the additional matter fields are included unlike the previous subsection and this is very similar to the 4d deconfinement \cite{Luty:1996cg}. Here we keep the ambiguities of the $U(1)$ charges. For even $N$, we should take even $K$.
For odd $N$, $K$ should be odd and we can take $K=1$. In that case, we don't need a $p_3$ field because the meson $p_1p_1$ cannot be constructed. Then the superpotential is simply $W=xp_1 p_2 $.

If we assume that the $Sp(N+K-2)$ theory first confines, then the low-energy theory is described by the mesons $A=xx, A_1:=xp_1, A_2:=p_1p_1$ and the Coulomb branch operators $Y_{sp}$ with the effective superpotential
\begin{align}
W= A_1 p_2 +A_2 p_3 -Y_{sp} \, \mathrm{Pf} \left(
    \begin{array}{cc}
      A& A_1  \\
      -A_1 & A_2 
    \end{array}
  \right).
\end{align}
The mesons $A_1$ and $A_2$ are massive due to the tree-level superpotential, then the effective superpotential is vanishing and plays no role. The important point is that the Coulomb branch operator $Y_{sp}$ is completely decoupled from other sectors. Thus we recover the electric $SU(N)$ gauge theory with $F$ fundamentals, $F'$ anti-fundamentals and one anti-symmetric without superpotentials.  

For the case with $F' =F+N-2$, the $SU(N)$ theory has the dual description \cite{Aharony:2013dha} with the (non-special) unitary gauge group. The dual is given by a $U(F+K-2)\times Sp(N+K-2)$ gauge theory. The superpotential is
\begin{align}
W=M_3 p_1 +p_1p_1 p_3 +M q \tilde{q}+M_1 q \tilde{p}_2 +M_2\tilde{x} \tilde{q} +M_3 \tilde{x} \tilde{p}_2 +Y_{SU} b \tilde{b}+\tilde{X}_+ +\tilde{X}_- ,
\end{align}
and the matter contents are summarized in Table \ref{DEC2}.
\begin{table}[H]\caption{Quantum numbers of the first dual theory} 
\begin{center}\scalebox{0.42}{
  \begin{tabular}{|c|c|c||c|c|c|c|c|c|c| } \hline
     &$U(F+K-2)$&$Sp(N+K-2)$& $SU(F)$ & $SU(F+N-2)$ &SU(K)& $U(1)_B$ & $U(1)_A$ & $U(1)_X$ & $U(1)_R$  \\ \hline
   $q$  & $\Box_1$  & $1$ & $\bar{{\tiny \yng(1)}}$ &1&1&$B-1$&$A-1$&$X+\frac{1}{2}$& $R$ \\
   $\tilde{q}$& $\bar{{\tiny \yng(1)}}_{-1}$ & $1$&  1& $\bar{{\tiny \yng(1)}}$ &1& $1-B$  &$-A-1$&$-X-\frac{1}{2}$& $2-R$ \\
$\tilde{x}$& $\Box_1$&$\Box$&1&1&1&$B$&$A$&$X$&$R$  \\
$p_1$&1&$\Box$&1&1&$\Box $&$B$&$A$&$X$&$R$ \\
$\tilde{p}_2$ &$\bar{ {\tiny \yng(1)}}_{-1}$ &1&1&1& $ {\tiny \yng(1)}$&0&0&0&0 \\
$p_3$&1&1&1&1& $\bar{{\tiny \yng(1,1)}}$&$-2B$&$-2A$&$-2X$&$2-2R$ \\
$b$&$1_{-(F+K-2)}$&1&1&1&1&$-B(F+K-2)+F$&$F-A(F+K-2)$&$-X(F+K-2) -\frac{1}{2}(F-N)$& $-R(F+K-2)$\\
$\tilde{b}$&$1_{F+K-2}$&1&1&1&1&$B(F-2)-N-F+2$&$A(F-2)+N+F-2$&$X(F-2)+\frac{1}{2}F-1$& $R(F-2)-2F+4$\\
$Y_{SU}$&1&1&1&1&1&$BK+N-2$&$AK-2F-N+2$&$XK-\frac{1}{2}N+1$&$RK+2F-2$ \\
$S$&1&1&1&1&1&$BK$&$AK$&$\frac{1}{2}N+XK$&$RK$ \\
$M=Q\tilde{Q}$&1&1&$\Box$&$\Box$&1&0&2&0&0 \\
$M_1=Qp_2$&1&1&$\Box$&1&$\bar{{\tiny \yng(1)}}$&$1-B$&$1-A$&$-\frac{1}{2}-X$&$2-R$ \\ 
$M_2=x \tilde{Q}$&1&$\Box$&1&$\Box$&1&$-1$&1&$\frac{1}{2}$&0 \\
$M_3=xp_2$&$1$&$\Box$&1&1&$\bar{{\tiny \yng(1)}}$&$-B$&$-A$&$-X$&$2-R$ \\ \hline 
$Y_{sp}^{\mathrm{mag}}$&$1_{-(F+K-2)}$&1&1&1&1&$-BK-BF+2B+F+N-2$&$-AK-AF+2A-F-N+2$&$-XF-XK+2X-\frac{1}{2}F-\frac{1}{2}N+1$&$-KR-FR+2R+2F-2$ \\ \hline 
  \end{tabular}}
  \end{center}\label{DEC2}
\end{table}
The advantage of this method is that the un-needed Coulomb branch is automatically decoupled then we don't worry about how to map these Coulomb branch operators and we can apply the conventional Seiberg or Aharony dualities without worrying about the UV realization of the terms such as $SY_{Sp}$. However, it is again difficult to see what degrees of freedom are decoupled from the dual theory. It would be interesting to see what degrees of freedom are decoupling by using the index calculations or the method presented in \cite{Lee:2016zud}. This direction will be left as the future work. In this first dual description, we expect that the Coulomb branch operator of the deconfined description $Y_{Sp}$ relates to the Coulomb branch operator of the $Sp(N+K-2)$ gauge group in a following way:
\begin{align}
Y_{Sp} =Y_{Sp}^{\mathrm{mag}}\tilde{b}.
\end{align}
This implies that the combination $Y_{Sp}^{\mathrm{mag}}\tilde{b}$ is completely decouples from the other moduli. This is highly non-trivial statement and it should be checked.

The superpotential gives the mass terms to the fields $M_3$ and $p_1$, then we can integrate out them. Then the theory reduces to
\begin{align}
W=(\tilde{x} \tilde{p}_2)(\tilde{x} \tilde{p}_2) p_3  +M q \tilde{q}+M_1 q \tilde{p}_2 +M_2\tilde{x} \tilde{q} +Y_{SU} b \tilde{b}+\tilde{X}_+ +\tilde{X}_- 
\end{align}
with the following matter contents.
\begin{table}[H]\caption{Quantum numbers of the first dual theory} 
\begin{center}\scalebox{0.42}{
  \begin{tabular}{|c|c|c||c|c|c|c|c|c|c| } \hline
     &$U(F+K-2)$&$Sp(N+K-2)$& $SU(F)$ & $SU(F+N-2)$ &SU(K)& $U(1)_B$ & $U(1)_A$ & $U(1)_X$ & $U(1)_R$  \\ \hline
   $q$  & $\Box_1$  & $1$ & $\bar{{\tiny \yng(1)}}$ &1&1&$B-1$&$A-1$&$X+\frac{1}{2}$& $R$ \\
   $\tilde{q}$& $\bar{{\tiny \yng(1)}}_{-1}$ & $1$&  1& $\bar{{\tiny \yng(1)}}$ &1& $1-B$  &$-A-1$&$-X-\frac{1}{2}$& $2-R$ \\
$\tilde{x}$& $\Box_1$&$\Box$&1&1&1&$B$&$A$&$X$&$R$  \\
$\tilde{p}_2$ &$\bar{ {\tiny \yng(1)}}_{-1}$ &1&1&1& $ {\tiny \yng(1)}$&0&0&0&0 \\
$p_3$&1&1&1&1& $\bar{{\tiny \yng(1,1)}}$&$-2B$&$-2A$&$-2X$&$2-2R$ \\
$b$&$1_{-(F+K-2)}$&1&1&1&1&$-B(F+K-2)+F$&$F-A(F+K-2)$&$-X(F+K-2) -\frac{1}{2}(F-N)$& $-R(F+K-2)$\\
$\tilde{b}$&$1_{F+K-2}$&1&1&1&1&$B(F-2)-N-F+2$&$A(F-2)+N+F-2$&$X(F-2)+\frac{1}{2}F-1$& $R(F-2)-2F+4$\\
$Y_{SU}$&1&1&1&1&1&$BK+N-2$&$AK-2F-N+2$&$XK-\frac{1}{2}N+1$&$RK+2F-2$ \\
$S$&1&1&1&1&1&$BK$&$AK$&$\frac{1}{2}N+XK$&$RK$ \\
$M=Q\tilde{Q}$&1&1&$\Box$&$\Box$&1&0&2&0&0 \\
$M_1=Qp_2$&1&1&$\Box$&1&$\bar{{\tiny \yng(1)}}$&$1-B$&$1-A$&$-\frac{1}{2}-X$&$2-R$ \\ 
$M_2=x \tilde{Q}$&1&$\Box$&1&$\Box$&1&$-1$&1&$\frac{1}{2}$&0 \\ \hline 
$Y_{sp}^{\mathrm{mag}}$&$1_{-(F+K-2)}$&1&1&1&1&$-BK-BF+2B+F+N-2$&$-AK-AF+2A-F-N+2$&$-XF-XK+2X-\frac{1}{2}F-\frac{1}{2}N+1$&$-KR-FR+2R+2F-2$ \\ \hline
  \end{tabular}}
  \end{center}\label{qcml1}
\end{table}

\section{Summary and Discussion}
In this paper we developed a deconfinement technique in 3d $\mathcal{N}=2$ supersymmetric gauge teories and investigated the dynamics of the product gauge group. This method is quite general and would be applicable to any two-index matters and general product gauge groups. We here concentrated on the anti-symmetric matters which were deconfined by the s-confining phase of the 3d $\mathcal{N}=2$ $Sp(2N)$ gauge theory. The mesonic operators are describing the Higgs branch and identified with the two-index matters we want. The difference between the 3d and 4d deconfinement methods is the presence of the Coulomb branch coordinates of the s-confining gauge theories. The unnecessary flat directions are lifted by introducing the additional singlets and by coupling them to the Coulomb branch coordinates as mass terms. In the ``s-confining $\times$ s-confining'' theories, these additional singlets are identified with the gauge singlets which consist of the two-index matters. For the theory with an anti-symmetric flavor, we have found the same massless direction as \cite{Csaki:2014cwa} but the form of the superpotential is slightly different. We have no satisfying reason to explain this discrepancy but we expect these two potentials describe the same physics. This would be left as the future work. For the analysis of the 3d $\mathcal{N}=2$ $SU(2)^N$ product gauge theory, we derived the exact superpotential which describes the low-energy 3d dynamics and this is highly consistent with the exact superpotential in 4d and with the Seiberg-Witten curve. 

The deconfined technique is usually not so useful because the dynamics of either $SU(N)$ or $Sp(2N')$ gauge groups is alway strong in the infrared region and we do not have the weakly coupled description. This situation is the same for the 3d deconfinement and what is even worse is that in 3d all the gauge theories become strong even if $U(1)$ gauge theories. However, while the 3d theories with two-index matters generally have the very complicated Coulomb moduli, the deconfined description have the very simple Coulomb branch since there are only the (bi-)fundamental matters in the deconfined theories. This is one of the advantages of the deconfinement in 3d. Furthermore, as we have shown, the 3d deconfinement with ``s-confining $\times$ s-confining'' structure gives the completely s-confining description with no gauge group for the $SU(N)$ gauge theory with anti-symmetric matters, which was absent in the 4d case with ``s-confining $\times$ s-confining''. It is very interesting to further investigate the deconfinement with ``s-confining $\times$ s-confining'' and find new s-confining descriptions of theories with two-index matters. It is also important to generalize the ``s-confining $\times$ s-confining'' structure to the cases with ``s-confining $\times$ quantum modified constraint'' or ``quantum modified constraint $\times$ quantum modified constraint'', which is easily performed by changing the matter contents. 

The difference of the deconfinement in 3d and 4d is that the 3d deconfinement and the duality transformation affect the structure of the Coulomb branch operators. The Coulomb branch of the 3d $\mathcal{N}=2$ $SU(N)$ gauge theory with anti-symmetric matters is very complicated and sometimes two coordinates are required to describe it \cite{Csaki:2014cwa, Amariti:2015kha}. In the deconfined description we have two Coulomb branch operators: One of them is a $SU(N)$ Coulomb branch and the other one comes from a $Sp(2N)$ gauge theory. Remember that in the deconfined description discussed in this paper, the $SU(N)$ part of the $SU \times Sp$ gauge theory is ``vector-like'', so we need only single operator for describing the $SU(N)$ Coulomb branch. These two Coulomb branches correctly explain the unlifted Coulomb branch of the 3d $\mathcal{N}=2$ $SU(N)$ gauge theory with anti-symmetric matters. When one first changes the dynamics of the $Sp$ part into the confined description, the Coulomb branch of the $SU(N)$ part becomes highly complicated as explained in \cite{Csaki:2014cwa, Amariti:2015kha}. If we first change the $SU(N)$ dynamics into the Seiberg dual or a confined one, the Coulomb branch of the $Sp$ part is modified because the fermions contributing to the $Sp$-monopole are mesons. When we further exchange the $Sp$ part to the dual description and obtain the second dual, the Coulomb branch of the $SU$ part becomes complicated since the duality transformation of the $Sp$ part introduces the two-index matters in the $SU(N)$ gauge group. In 4d, we can think of the $SU$ gauge group as a flavor symmetry under the $Sp$ Seiberg duality transformation and only the matter contents change. The modification of the Coulomb branch under the duality sequence is peculiar to the 3d deconfinement.

We showed the alternative way of the deconfinement, namely decoupling the Coulomb moduli. In this method we do not lift the unnecessary Coulomb branch but only decouple it from the other sector by introducing the additional matters. This is successful but it is difficult to find the decoupled direction in the dual theories. In order to deal with this problem, it would be preferable to find the decoupled degrees of freedom on the dual side, for example, by using the argument in \cite{Lee:2016zud}, where the quantum dimensions of the operators are investigated by means of the F-maximization and the decoupled directions are studied. We can use the same technique in our example.

In this paper we checked the validity of the deconfinement with ``s-confining $\times$ s-confining'' by the correspondence of the moduli space and the comparison with the known dualities. Although this check is quite strong, it is worth performing the another checks. For example, we can compare the (superconformal) indices of the electric and magnetic theories by using the localization method \cite{Pestun:2007rz, Kapustin:2009kz, Kapustin:2010xq, Hama:2010av, Imamura:2011su}. Furthermore it is important to find the correct $U(1)_R$ charge realized in the infrared comformal fixed point, which would be possible if we use the F-maximization. If the duality is correct, the infrared $U(1)_R$ charges on the electric and magnetic sides should match. This would be a non-trivial check of the duality.

It is also possible to apply the 3d deconfinement technique to (supersymmetric) Chern-Simons matter theories with two index matters. This is a simple generalization and all the story which we gave in this paper can be applied. In Section 5, we studied the deconfinement of the 3d $\mathcal{N}=2$ $SU(N)$ gauge theory with one anti-symmetric, $F$ fundamentals and $F'=F+2N-2$ anti-fundamentals. The deconfined theory is a vector-like theory and its dual is given by the $U(F-N) \times Sp(N+K-2)$ gauge theory. It is interesting to consider the chiral version. The Seiberg duality of the chiral gauge theory is considered in \cite{Benini:2011mf, Aharony:2014uya, Aharony:2013dha} and we can use it for finding the duality of the 3d $\mathcal{N}=2$ $SU(N)$ gauge theory with one anti-symmetric, $F$ fundamentals and general $F'$ anti-fundamentals. It is also possible and interesting to study the deconfinement of adjoint matters and symmetric matters. These are described in terms of the s-confining phases of the $(S)U(N)$ and $(S)O(N)$ gauge theories respectively.

Now we have the 3d and 4d deconfinement methods. Then it is quite intriguing to understand the relation between them. It would be possible to dimensionally reduce the 4d deconfinement description into 3d one because the deconfined description only contains the fundamental matters, so the reduction and the deformation required for obtaining the 3d duality would be the same as \cite{Aharony:2013dha}. The procedure is very simple: First, we put the 4d theory on a circle and take into account the effect of the compactification. The effect mainly comes from the KK-monopoles (twisted instantons) \cite{Lee:1997vp}, which generate the non-perturbative superpotentials. The same step is also carried out for the dual theory. In this way we can obtain the 3d or $\mathbb{R}^3 \times \mathbb{S}^1$ duality with some superpotentials. In order to obtain the duality without the superpotentials, we need to further deform the theory by giving real masses and taking the low-energy limit. On the dual side we need to find the same deformation as the electric side. 
However since the global $U(1)$ charge assignment of the matter contents is complicated in the deconfinement theory, it is difficult to find where the low-energy limit should be taken in the dual theory.
Thus it is very interesting problem to find the connection of the 3d and 4d deconfinement methods.

In Sec.2, we discussed the dimensional reduction of the special class of the $\mathcal{N}=1$ Seiberg-Witten curve where the 4d theory only has the Coulomb phase with rank $1$. But one can generally write down the Seiberg-Witten curves for theories with also the Higgs branch. The dimensional reduction would well work for such theories and one can obtain the effective 3d superpotentials including the Higgs and Coulomb moduli from the Seiberg-Witten curve. We will briefly show the story of this direction. The Seiberg-Witten curve generically contains the mass or coupling dependence for the matter chiral superfields and in this sense the Higgs branch is integrated out. In the dimensional reduction of the theory whose dynamics on the Coulomb branch is described by Seiberg-Witten curve, we can interpret the variables $x,y$ describing the curve as the dynamical fields. Integrating in the matter contribution into the superpotential by inverse Legendre transformations we obtain the correct description of the Coulomb and Higgs branches. To be more specific, let us consider the 4d $\mathcal{N}=2$ $SU(2)$ gauge theory with $N_f=1$ fundamental flavors \cite{Seiberg:1994aj}. The corresponding Seiberg-Witten curve is 
\begin{align}
y^2 = x^2(x-u) +2mx +1, 
\end{align}
where $m$ is a mass term for the hypermultiplet and $u:=\mathrm{tr} \phi^2$.
By changing the variables as $x-u =:v$, compactifying the theory on a circle and introducing the soft SUSY breaking term to the adjoint chiral superfield, we obtain the superpotential
\begin{align}
W=\lambda (y^2 - x^2v -2mx -1) +\epsilon (x \eta -v),
\end{align}
where $\epsilon$ is a SUSY breaking parameter, $\eta$ is a 4d dynamical scale and $\lambda$ is a Langrange multiplier to impose the defining equation of the Seiberg-Witten curve. Integrating out $\lambda, y$ and $v$, we find the low-energy superpotential
\begin{align}
W=\epsilon x \eta +\epsilon \frac{1-2mx}{x^2}.
\end{align}
Note that since the effective superpotential has the linear dependence on $m$, then Legendre transforming $m$ to the meson field $M$ gives the identification $M \sim \frac{2\epsilon}{x}$. If we rescale $x$ as $Y:=x/(2 \epsilon)$ and interpret this as the Coulomb branch operator, we can find the quantum modified constraint $MY =1$, which is found in \cite{Aharony:1997bx}. In this rescale and by taking the 3d limit where $\eta \rightarrow 0$, the first and second terms in the superpotential are vanishing. This analysis is straightforwardly generalized to $N_f=2$ \cite{Seiberg:1994aj}, in which the Seiberg-Witten curve takes
\begin{align}
y^2 =(x^2-1)v +2m_1 m_2 x-(m_1^2+m_2^2)
\end{align}
in terms of the rescaled variables. We can integrate in the meson fields $M_{12}$ and $M_{34}$ and find
\begin{align}
W= -\frac{M_{12}M_{34} x}{2\epsilon} +\cdots = - M_{12}M_{34} Y+\cdots,
\end{align}
where the ellipses are terms vanishing in the limit with $\epsilon \rightarrow \infty$ and $\eta \rightarrow 0$. This result is beautifully consistent with the effective superpotential $W=-Y \mathrm{Pf} M$. Notice that since the above derivation relies on the integrating-in method, we may in general include other terms and in many cases we are missing new massless degrees of freedom appearing in the origin of the moduli space. The similar analyses are found in \cite{Boels:2003fh,Boels:2004ua,Boels:2003at,Dorey:1999sj} and it would be possible to generalize the analysis to more complicated $\mathcal{N}=1$ Seiberg-Witten curves. It is also worth investigating the 3d $\mathcal{N}=2$ $SU(N) \times SU(N)$ theory with bi-fundamental matters since the 4d version of this theory has Coulomb phases and we will be able to find the interesting physics in 3d.

\section*{Acknowledgments}

 I would like to thank Ofer Aharony and Ken Intriligator for useful discussions. I am also grateful to Ashoke
Sen for helpful comments.

\bibliographystyle{ieeetr}
\bibliography{3dref}

\end{document}